\documentclass[aps,prx,reprint,twocolumn, amsmath,amssymb,amsfonts,footinbib,
superscriptaddress,longbibliography,
notitlepage]{revtex4-2}

\AtBeginDocument{\usepackage{booktabs}}               
\makeatletter
\g@addto@macro\bfseries{\boldmath}
\makeatother

\usepackage[varg]{txfonts}
\usepackage[T1]{fontenc}
\usepackage[utf8]{inputenc}

\usepackage[varg]{txfonts}
\usepackage{hyperref}
\usepackage{amsmath}
\usepackage{color}
\usepackage{graphicx}
\usepackage[percent]{overpic}
\usepackage{mathrsfs}
\usepackage{bm}
\usepackage{braket}
\usepackage{mathtools}
\usepackage{bbold}
\usepackage{enumitem}
\usepackage{xcolor}
\usepackage[normalem]{ulem}

\usepackage{makecell}

\definecolor{cred}{RGB}{228,26,28}
\definecolor{cblue}{RGB}{8,48,107}
\definecolor{cgreen}{RGB}{77,175,74}
\definecolor{cgray}{RGB}{150,150,150}
\definecolor{clgray}{RGB}{200,200,200}
\definecolor{cpurple}{RGB}{152,78,163}
\definecolor{corange}{RGB}{255,127,0}
\definecolor{cgold}{RGB}{230,171,2}

\newcommand{\h}[1]{{#1}^{\dagger}}

\newcommand{\sgn}{{\rm sgn}}

\newcommand{\tr}{{\rm Tr}}

\newcommand{\subref}[2]{\ref{#1}\hyperref[#1]{#2}}

\renewcommand{\vec}[1]{\boldsymbol{#1}}
\renewcommand{\l}[1]{{#1}^{\phantom{\dagger}}} 
\newcommand{\mat}[1]{\vec{#1}}

\newcommand{\vhat}[1]{\vec{\hat{#1}}}

\definecolor{cred}{RGB}{188,55,84}
\hypersetup{colorlinks=true,linkcolor=blue,citecolor=blue,urlcolor=blue}

\graphicspath{{./figs/}}
\begin{document}

\title{Magnon topology driven by altermagnetism}

\author{Subhankar Khatua}
\affiliation{Institute for Theoretical Solid State Physics, IFW Dresden, 01069 Dresden, Germany}
\affiliation{Würzburg-Dresden Cluster of Excellence ct.qmat, 01069 Dresden, Germany}
\author{Volodymyr P. Kravchuk}
\affiliation{Institute for Theoretical Solid State Physics, IFW Dresden, 01069 Dresden, Germany}
\affiliation{Würzburg-Dresden Cluster of Excellence ct.qmat, 01069 Dresden, Germany}
\affiliation{Bogolyubov Institute for Theoretical Physics of the National Academy of Sciences of Ukraine, 03143 Kyiv, Ukraine}
\author{Kostiantyn V. Yershov}
\affiliation{Institute for Theoretical Solid State Physics, IFW Dresden, 01069 Dresden, Germany}
\affiliation{Bogolyubov Institute for Theoretical Physics of the National Academy of Sciences of Ukraine, 03143 Kyiv, Ukraine}
\author{Jeroen van den Brink}
\affiliation{Institute for Theoretical Solid State Physics, IFW Dresden, 01069 Dresden, Germany}
\affiliation{Würzburg-Dresden Cluster of Excellence ct.qmat, 01069 Dresden, Germany}
\affiliation{Institute of Theoretical Physics,
Technische Universit\"at Dresden, 01062 Dresden, Germany}

\begin{abstract}
Altermagnets present a class of fully compensated collinear magnetic order, where the two sublattices are not related merely by time-reversal combined with lattice translation or inversion, but require an additional lattice rotation. This distinctive symmetry leads to a characteristic splitting of the magnon bands; however the splitting is only partial -- residual degeneracies persist along certain lines in the Brillouin zone as a consequence of the underlying altermagnetic rotation. We consider a two-dimensional $d$-wave altermagnetic spin model on the checkerboard lattice and introduce additional interactions such as an external magnetic field and Dzyaloshinskii-Moriya interactions, that lift these degeneracies. The resulting magnon bands become fully gapped and acquire non-trivial topology, characterized by nonzero Chern numbers. We demonstrate the crucial role of altermagnetism for the generation of the Berry curvature. As a direct consequence of the topological magnons, we find finite thermal Hall conductivity $\kappa_{xy}$, which exhibits a characteristic low-temperature scaling, $\kappa_{xy}\propto T^4$. Moreover, $\kappa_{xy}$ changes sign under reversal of the magnetic field, exhibiting a sharp jump across zero field at low temperatures. We also demonstrate topologically protected chiral edge modes in a finite strip geometry. 

\end{abstract}

\date{\today}

\maketitle
\section{Introduction}
Topological states of matter constitute a highly active area of research in modern condensed matter physics~\cite{Hasan2010,Qi2011,Bernevig2013,Senthil2015,Karzig2015,Wen2017,Sato2017,Armitage2018,Burkov2018,Ozawa2019,Cooper2019,McClarty2022,Okuma2023,Corbae2023}. 
One of the most extensively studied directions in this context is electronic topological band insulators~\cite{Hasan2010,Qi2011,Bernevig2013,Wen2017}.  
A hallmark of such insulators is the presence of gapped bulk bands characterized by certain topological invariants, and the gapless edge or surface states -- leading to non-trivial transport phenomena such as quantized Hall conductance. 
While the basic ideas of topological band structures and subsequent theoretical as well as experimental progress have occurred primarily in electronic (fermionic) systems, many of these concepts have also been extended to bosonic systems~\cite{Karzig2015,Ozawa2019,McClarty2022,Romhanyi2015}, such as magnons in magnetic Mott insulators.

Magnons are quantized collective spin-wave excitations in magnetically long-range ordered phases and are ubiquitous in insulating magnetic materials~\cite{kittel2005,Fazekas1999}. 
Since magnons are quasiparticles described by Bloch wavefunctions in crystalline solids, in analogy with electronic bands, magnon bands in momentum space can also be associated with topological invariants~\cite{McClarty2022}. 
Geometric quantities such as the Berry phase~\cite{Berry1984}, Berry connection and curvature can be defined for a magnon band.  
In particular, in two-dimensional systems, one can define a topological invariant known as the Chern number, expressed in terms of the Berry curvature, and a non-zero Chern number signals non-trivial band topology~\cite{McClarty2022}.

Unlike electrons, magnons are charge-neutral and thus are not affected by Lorentz forces. 
However, they do transport heat, and under a temperature gradient, a magnonic heat current can be induced. 
Consequently, topological magnons can give rise to thermal Hall effect~\cite{Katsura2010,Matsumoto2014}, as observed experimentally in pyrochlore ferromagnets~\cite{Onose2010,Ideue2012}, kagome ferromagnets~\cite{Chisnell2015,Hirschberger2015}, the Kitaev material $\alpha$-RuCl$_3$ in the field-polarized phase~\cite{Czajka2023}. 
Due to their bosonic nature, the thermal Hall response of topological magnons is not quantized, in contrast to its electronic counterpart.    
Beyond their fundamental importance in magnetism, topological magnons hold great potential for applications in spintronic and magnonic devices, owing to their tunability via magnetic fields and the absence of Joule heating~\cite{Chumak2015,Wang2018}.

Topological magnons have been primarily investigated in collinear ferromagnets (FMs) with Dzyaloshinskii-Moriya interaction (DMI)~\cite{Dzyaloshinsky1958,Moriya1960}, where DMI plays a crucial role in generating nontrivial topology of the magnon bands~\cite{Zhang2013,Mook2014a,Mook2014b,Cao2015,Owerre2016a,Owerre2016b,Mook2016,Kim2016,Chen2018,Seshadri2018,Malki2019,Pires2021,Zhuo2021}. 
However, such topological features are not as prevalent in antiferromagnets (AFMs) as in FMs. 
In fully compensated collinear Heisenberg AFMs, two magnon branches are completely degenerate due to symmetries such as time-reversal combined with spatial inversion or translation.  
The presence of any band degeneracies hinders the conventional computation of the Berry curvature and Chern number~\cite{Bhowmick2020}.  
Nonetheless, a number of antiferromagnetic models have been proposed in which DMI and external magnetic fields stabilize non-collinear magnetic orderings and give rise to gapped out topological magnon bands~\cite{Owerre2016c,Owerre2017,Owerre2017_kag,Laurell2017,Laurell2018,Kawano19,Bhowmick2020,Neumann2022,Chen2023}.

Beyond standard FMs and AFMs, there is a recently identified class of magnetic order known as \emph{altermagnet} (AM) which is currently attracting significant attention~\cite{Smejkal2022a,Smejkal2022b}. 
AMs exhibit fully compensated N\'eel magnetic order, yet are symmetry-wise distinct from AFMs. 
In collinear AFMs, the two magnetic sublattices are related by  time-reversal (spin-flip) combined with spatial inversion or translation. However, in AMs, connecting two sublattices additionally requires a spatial rotation. 
As a result of this added symmetry operation, the magnon branches in AMs are \emph{not} fully degenerate, as they are in AFMs without spin-orbit coupling effects, but are instead related to one another by the same rotation~\cite{Yershov2024,Consoli2025}.
This lifting of degeneracy by (longer range) exchange interactions endows AMs with properties akin to ferromagnets such as the ability to support spin currents~\cite{Smejkal2022b}. 
While sharing certain features with both FMs and AFMs, AMs remain largely unexplored in the context of topological magnons~\cite{zhang2025}. 
This naturally raises the question of whether altermagnetism can induce topological magnons. 
Notably, since the band degeneracy is already partially lifted in AMs, the addition of perturbative interactions can fully remove residual degeneracies, allowing the exploration of gapped magnon band topology.

In this article, we investigate topological magnons in a minimal two-dimensional altermagnetic spin model -- the checkerboard lattice spin model~\cite{Canals2002, Yershov2024,Consoli2025} in the presence of an out-of-plane uniform Zeeman field and DMI. 
The altermagnetic checkerboard lattice spin model consists of two exchange couplings -- antiferromagnetic Heisenberg coupling between nearest-neighbors and ferromagnetic or antiferromagnetic Heisenberg coupling between next-nearest-neighbors along the diagonals of the checkerboard plaquettes. 
The latter is crucial for realizing altermagnetism in this model~\cite{Yershov2024,Consoli2025}; in its absence, this model reduces to the nearest-neighbor square-lattice AFM.  
Accordingly, we refer to the diagonal coupling henceforth as the \emph{AM coupling}.  
This checkerboard lattice model stabilizes a fully compensated collinear altermagnetic ground state, with its two magnon energy-branches displaying $d$-wave-like splitting, remaining degenerate only along certain lines in the Brillouin zone~\cite{Yershov2024,Consoli2025}. 
We find that introducing a uniform out-of-plane Zeeman field in this model stabilizes a spin-flop phase: spins cant out of the lattice-plane towards the field, and moreover, this canted state remains the ground state even upon adding the specific DMI that we consider. 
Using linear spin-wave theory, we analytically show that when all three ingredients -- AM coupling, Zeeman field, and DMI -- are present, two magnon branches open a gap with respect to each other. 
Interestingly, the energy gap between the bands at the mid-points of the Brillouin zone edges is determined solely by the AM coupling.  
To reveal the topological character of the magnon bands, we calculate the Berry connection, Berry curvature, and find the Chern number to be non-zero, namely $\pm1$, and opposite for the upper and lower branches. 
Furthermore, reversing the sign of any of the three couplings -- Zeeman field, AM coupling, or DMI, reverses the sign of the Berry curvature and hence the Chern number.  
In the limit of vanishing AM coupling, the Berry curvature becomes concentrated near the midpoints of the Brillouin zone edges, forming elliptical regions whose principal axes are set by the DMI and Zeeman field, respectively. 
As a consequence of the topological magnons, we compute thermal Hall conductivity, $\kappa_{xy}$, under a temperature gradient.
At low temperatures, $\kappa_{xy}$ is found to scale with temperature $T$ as $\kappa_{xy}\propto T^4$. 
Moreover, $\kappa_{xy}$ changes sign upon flipping the Zeeman field, displaying a sharp jump across zero field at low temperatures.
Finally, as another non-trivial manifestation of magnon band topology, we also demonstrate the existence of chiral edge states on a finite strip geometry.

\section{Model}
\label{sec.model}
We consider the following spin model on the checkerboard lattice described by the Hamiltonian
%%%%%%%%%%%%%%%%%%%%%%%%%%%%%%%%%%
\begin{figure}
\centering
\includegraphics[width=0.95\columnwidth]{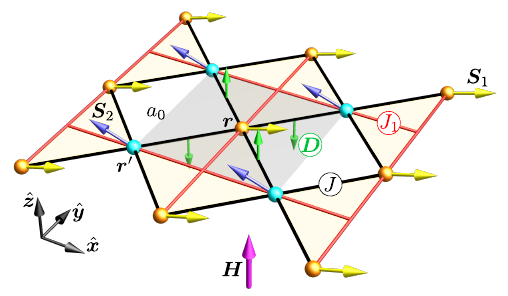}
\caption{Checkerboard lattice spin model in the presence of a Zeeman field and DMI: The lattice consists of two interpenetrating square sublattices-1 and 2 whose site positions are denoted by $\vec{r}$ and $\vec{r}'$, respectively, each with lattice constant $a_0$. Spin-$S$ operators reside at every site of the resulting checkerboard lattice. Nearest-neighbor spins across the two sublattices interact via an antiferromagnetic Heisenberg exchange ($J > 0$, black bonds), while next-nearest-neighbor spins within each sublattice along the diagonals of the checkerboard plaquettes interact either ferromagnetically or antiferromagnetically ($J_1$, red bonds). Out-of-plane Dzyaloshinskii-Moriya vectors $\vec{D}$ are indicated by green arrows on the nearest-neighbor bonds, and $\vec{H}$ shows the direction of the uniform Zeeman field. The light gray square of side $a_0$ marks the magnetic unit cell of the checkerboard lattice. Arrows at the lattice sites represent the canted classical ground-state configuration of the model.} 
\label{fig.model}
\end{figure}
%%%%%%%%%%%%%%%%%%%%%%%%%%%%%%%%%%
\begin{align}
    {\cal H} &= J\sum_{\langle\vec{r}\vec{r}'\rangle}\vec{S}_{\vec{r},\,1}\cdot\vec{S}_{\vec{r}',\,2} + J_1\sum_{\vec{r}}\vec{S}_{\vec{r},\,1}\cdot\vec{S}_{\vec{r}\, + \,a_0\vhat{y},\,1} \nonumber\\ 
    &+J_1\sum_{\vec{r}'}\vec{S}_{\vec{r}',\,2}\cdot\vec{S}_{\vec{r}'\,+\,a_0\vhat{x},\,2}- g\mu_\textsc{b}\mu_0\vec{H}\cdot\bigg[\!\sum_{\vec{r}}\vec{S}_{\vec{r},\,1} + \sum_{\vec{r}'}\vec{S}_{\vec{r}',\,2}\bigg]\nonumber\\
    &+ D\vhat{z}\cdot\!\!\sum_{\vec{r}}\vec{S}_{\vec{r},\,1}\times \Big[\vec{S}_{\vec{r}\,+\, \vec{\delta}_\searrow,\,2} + \vec{S}_{\vec{r}\,-\, \vec{\delta}_\searrow,\,2}-\vec{S}_{\vec{r}\,+\, \vec{\delta}_\nearrow,\,2}-\vec{S}_{\vec{r}\,-\, \vec{\delta}_\nearrow,\,2}\Big],
    \label{eq.model-Ham}
\end{align}
where $\cramped{\vec{r} =  a_0(n_x\vhat{x} + n_y\vhat{y})}$ and $\cramped{\vec{r}' = \vec{r}-\frac{a_0}{2}(\vhat{x} + \vhat{y})}$ denote the positions of sites on two interpenetrating square sublattices, each with lattice constant $a_0$, and $\cramped{n_x, n_y \in \mathbb{Z}}$, together forming a checkerboard lattice as illustrated in Fig.~\ref{fig.model}. 
$\cramped{\vec{S}_{\vec{r},\,1}\equiv \big(S_{\vec{r},\,1}^x,S_{\vec{r},\,1}^y,S_{\vec{r},\,1}^z\big)}$ and $\cramped{\vec{S}_{\vec{r}',\,2}\equiv \big(S_{\vec{r}',\,2}^x,S_{\vec{r}',\,2}^y,S_{\vec{r}',\,2}^z\big)}$ represent dimensionless spin-$S$ operators at $\vec{r}^{\rm th}$ site of sublattice-$1$ and ${\vec{r}'}^{\rm th}$ site of sublattice-$2$, respectively. 
The first term in the Hamiltonian corresponds to antiferromagnetic Heisenberg exchange ($\cramped{J > 0}$) between nearest-neighbor spins across sublattices, while the second and third terms represent Heisenberg exchange, either ferromagnetic $(J_1<0)$ or antiferromagnetic $(J_1>0)$, 
between next-nearest-neighbor spins within each sublattice along the checkerboard diagonals [see Fig.~\ref{fig.model}]. 
The fourth term describes the Zeeman coupling of a uniform external magnetic field $\vec{H} = H\vhat{z}$ to the magnetic moments associated with the spins: $\vec{\mu}_{\vec{r},1} \equiv g\mu_\textsc{b}\vec{S}_{\vec{r},1}$ and $\vec{\mu}_{\vec{r}',2} \equiv g\mu_\textsc{b}\vec{S}_{\vec{r}',2}$, where $g$ is the Land\'e $g$-factor of the spins and $\mu_\textsc{b}$ is the Bohr magneton.   
Finally, the last term corresponds to DMI between nearest-neighbor spins, with $\vec{\delta}_\searrow = \frac{a_0}{2}(\vhat{x}\,-\,\vhat{y})$, $\vec{\delta}_\nearrow = \frac{a_0}{2}(\vhat{x}\,+\,\vhat{y})$, and the out-of-plane Dzyaloshinskii-Moriya vectors $\vec{D}$ as shown in Fig.~\ref{fig.model}. Such a DMI is allowed by Moriya’s rules~\cite{Moriya1960}: the midpoints of the nearest-neighbor bonds lack inversion symmetry due to the presence of the $J_1$ bonds, and since the lattice plane of an ideal checkerboard lattice serves as a mirror plane, the corresponding Dzyaloshinskii–Moriya vectors must be perpendicular to it, i.e., out of plane.

There are two interesting symmetries of the full model [Eq.\eqref{eq.model-Ham}].
First is the ${\rm U}(1)$ spin-rotation symmetry about the $\vhat{z}$ axis, and the
second is the $C_4$ lattice-rotation about the $\vhat{z}$ axis passing through the center of any checkerboard plaquette, i.e., the center of any $J_1$ bond.
The latter interchanges the two square-sublattice sites of the checkerboard lattice.
As seen in Fig.~\ref{fig.model}, this transformation moves the lattice sites in such a way that the summed contributions from each term -- the $J$ bonds, the $J_1$ bonds, the Zeeman term, and the DMI term -- remain individually invariant. 
The invariance of the DMI term may not be immediately obvious, since under $C_4$ rotation, each bond maps to a different bond on which the Dzyaloshinskii-Moriya vector has the opposite sign.
However, since the $C_4$ operation exchanges the two sublattices, the order of the spin-spin cross-product in the DMI term reverses, introducing a compensating minus sign.
As a result, the total DMI contribution also remains invariant under the $C_4$ lattice-rotation. 
Additionally, the model exhibits inversion symmetry about the center of any $J_1$ bond; however, this is not an independent symmetry, as it is equivalent to applying two successive aforementioned $C_4$ lattice-rotation.

\section{Classical ground states}
Classically, the spins are three-component vectors of length $S$. 
In the absence of both the Zeeman field and DMI (i.e., $H = D = 0$), the model exhibits two distinct limits. 
First, when $J_1 = 0$, the classical ground state is a N\'eel state that can point in any arbitrary direction.  
Notably, such a N\'eel ordering is \emph{antiferromagnetic}, since the two magnetic sublattices are related by a spin-flip combined with a unit square-lattice-translation (i.e., translation by a black bond in Fig.~\ref{fig.model}). 
The other limit corresponds to $J_1\neq 0$. 
As mentioned in Sec.~\ref{sec.model}, $J_1$ can be either positive or negative.
When $\cramped{J_1 <0}$, it is straightforward to verify that the N\'eel state minimizes both $J$ and $J_1$ terms simultaneously. 
For $\cramped{J_1>0}$, although the $J_1$ bonds are not obviously minimized in the N\'eel configuration, the Luttinger-Tisza method shows that the N\'eel state still remains the classical ground state as long as $\cramped{J_1<J}$~\cite{Canals2002}.  
Importantly, the N\'eel ground state for $J_1=0$ is symmetry-wise distinct from that for $J_1\neq 0$. 
While the former is an antiferromagnetic ordering as argued above, the latter is not.  
The latter is rather an \emph{altermagnetic ordering}; the two magnetic sublattices are related by a spin-flip combined with a $C_4$ lattice-rotation about the $\vhat{z}$ axis passing through the center of any $J_1$ bond~\cite{Yershov2024,Consoli2025}. 
The checkerboard diagonal interaction $J_1$ thus introduces altermagnetism in the model, and referred to as the \emph{AM coupling} as mentioned in the Introduction.    
 
To find the ground states when $\cramped{D \neq 0}$ and $\cramped{H \neq 0}$, we consider a two-sublattice ground state parametrized by 
\begin{align}
\vec{S}_{\vec{r},1}&=S\left(\sin\theta_1\cos\phi_1\,\vhat{x}+\sin\theta_1\sin\phi_1\,\vhat{y}+ \cos\theta_1\,\vhat{z}\right),\nonumber\\
\vec{S}_{\vec{r}',2}&=S\left(\sin\theta_2\cos\phi_2\,\vhat{x}+\sin\theta_2\sin\phi_2\,\vhat{y}+ \cos\theta_2\,\vhat{z}\right),
\end{align}
where $\{\vhat{x},\vhat{y},\vhat{z}\}$ constitutes a right-handed \emph{laboratory} coordinate frame [see Fig.~\ref{fig.model}], and $\theta_\alpha$ and $\phi_\alpha$ with $\alpha=1,\,2$ are the polar and azimuthal angles specifying the classical spin-direction on sublattice-$\alpha$.  
Using these spin vectors, and assuming the spatially uniform magnetization of each of the sublatices, we compute the energy from Eq.~\eqref{eq.model-Ham} and find the following: since the AM coupling ($J_1$) terms involve spins only within the same sublattice, they do not depend on $\theta_\alpha$ or $\phi_\alpha$. 
Moreover, the total DMI for the four nearest-neighbor bonds of any sublattice sums to zero because the total Dzyaloshinskii-Moriya vector of a magnetic unit cell vanishes (see Fig.~\ref{fig.model}).
Consequently, the ground state energy does not depend on DMI. 
Therefore, determining $\theta_\alpha$ and $\phi_\alpha$ depends only on $J$ and the Zeeman field, reducing the problem to finding the ground-state configuration of a nearest-neighbor Heisenberg AFM on the square lattice in the presence of a uniform Zeeman field~\cite{Zhitomirsky1998}. 
The ground state undergoes a coplanar spin-flop phase where the two sublattices cant out of the lattice plane towards the field direction by an angle $\xi = \sin^{-1}\left(\frac{g\mu_\textsc{b}\mu_0H}{8JS}\right)$ with the lattice plane. 
This yields $\theta_1 = \theta_2 = \pi/2 - \xi$ and $\phi_1 - \phi_2 = \pi$. 
At $H = 0$, we have $\xi = 0$ corresponding to a N\'eel state in the lattice plane. 
As $H$ increases, two sublattices progressively cant towards the field direction. 
At the saturation field, ${H}_s = \frac{8JS}{g\mu_\textsc{b}\mu_0}$, we have $\xi = \pi/2$, and the system reaches a fully polarized configuration along the field direction. 
As a general reference ground state, we may choose $\phi_1 = 0$ and $\phi_2 = \pi$, such that the ground state is given by
\begin{eqnarray}
\label{eq.ground-state}
    \vec{S}_{\vec{r},1}\equiv\vec{S}_1 &=& S(\cos\xi\,\vhat{x} + \sin\xi\,\vhat{z}),\nonumber\\
    \vec{S}_{\vec{r}',2}\equiv\vec{S}_2 &=& S(-\cos\xi\,\vhat{x} + \sin\xi\,\vhat{z}).
    \label{eq.cgs}
\end{eqnarray}
A schematic illustration of this canted ground state is shown in Fig.~\ref{fig.model}. 
With these classical ground states we will next perform linear spin-wave analysis in order to obtain the magnon energy spectrum. 

\section{Spin-wave theory and magnon bands}
\label{sec.swt-magnon-bands}
We introduce small quantum fluctuations around the classical long-range magnetic order described in Eq.~\eqref{eq.cgs} by bosonizing the Hamiltonian [Eq.~\eqref{eq.model-Ham}] using the Holstein-Primakoff transformation~\cite{Holstein1940}. Retaining terms up to quadratic order in the bosonic operators, we obtain the linear spin-wave Hamiltonian
\begin{equation}
     \hspace{-0.1 cm}{\cal H}_2 \!=\!\! \sum_{\alpha\beta,\vec{q}}\left[{A}^{\alpha\beta}_{\vec{q}}\h{a}_{\vec{q},\alpha}\l{a}_{\vec{q},\beta} +
          \frac{1}{2} \left({B}^{\alpha\beta}_{\vec{q}}\h{a}_{\vec{q},\alpha}\h{a}_{-\vec{q},\beta} +
          \bar{B}^{\alpha\beta}_{\vec{q}}\l{a}_{-\vec{q},\beta}\l{a}_{\vec{q},\alpha}\right)
          \right],
          \label{eq.lsw-Ham}
\end{equation}
where $\h{a}_{\vec{q},\alpha}$ and $\l{a}_{\vec{q},\alpha}$ denote the bosonic creation and annihilation operators, respectively, at wave vector $\vec{q}$ on sublattice-$\alpha$. 
The coefficients are given by
\begin{eqnarray}\label{eq:H-components}
A_{\vec{q}}^{11} &=&  E_0\left(1-\frac{J_1}{J}\sin^2 \frac{q_ya_0}{2}\right),\,A_{\vec{q}}^{22} =  E_0\left(1-\frac{J_1}{J}\sin^2\frac{q_xa_0}{2}\right),\nonumber\\
A_{\vec{q}}^{12} &=&\bar{A}_{\vec{q}}^{21}= -E_0\sin\xi\left(\sin\xi\cos\frac{q_xa_0}{2}\cos\frac{q_ya_0}{2}\right.\nonumber\\
&&\hspace{2.5 cm}\left.-\,i\,\frac{D}{J}\sin\frac{q_xa_0}{2}\sin\frac{q_ya_0}{2}\right),\nonumber\\
B_{\vec{q}}^{11} &=& B_{\vec{q}}^{22} = 0,\,
B_{\vec{q}}^{12} = B_{\vec{q}}^{21}= E_0\cos^2\xi\cos\frac{q_xa_0}{2}\cos\frac{q_ya_0}{2}.
\label{eq.swH-coeff}
\end{eqnarray}
where $E_0 \equiv 4JS$, and $\bar{A}_{\vec{q}}^{21}$ denotes the complex conjugate of $A_{\vec{q}}^{21}$. 
A detailed derivation of the linear spin-wave Hamiltonian and its analytical diagonalization via a para-unitary transformation used to obtain the eigenvalues and eigenvectors is presented in Appendix~\ref{app.bosonic-sw}. 
%%%%%%%%%%%%%%%%%%%%%%%%%%%%%%%%%%
\begin{figure}[b]
\includegraphics[width=\columnwidth]{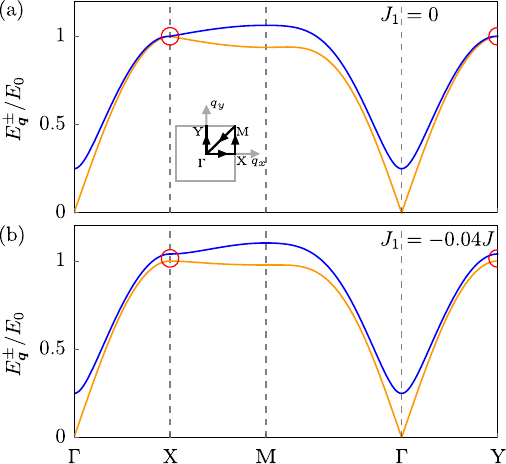}
\caption{Linear spin-wave spectrum [Eq.~\eqref{eq:Epm}] with the Zeeman field $H = H_s/8$ and DMI constant $D = 0.5J$ for a path in the first Brillouin zone ${\rm \Gamma}\,(0,0)\rightarrow {\rm X}\,(\pi/a_0,0)\rightarrow {\rm M}\,(\pi/a_0,\pi/a_0)\rightarrow {\rm \Gamma}\,(0,0)\rightarrow {\rm Y}\,(0,\pi/a_0)$. (a) Spectrum with the AM coupling $J_1 = 0$ shows degeneracy at X and Y, marked by the red circle and (b) with $J_1 = -0.04J$, a gap opens up at X and Y.}
\label{fig.specs}
\end{figure}
%%%%%%%%%%%%%%%%%%%%%%%%%%%%%%%%%%
The energy eigenvalues of the magnon branches are given by
\begin{eqnarray}\label{eq:Epm}
\hspace{-0.2 cm}\hbar\omega^{\pm}_{\vec{q}} =
 \frac{1}{\sqrt{2}}\sqrt{\left(A^{11}_{\vec{q}}\right)^2 + \left(A^{22}_{\vec{q}}\right)^2 - 2\left(B^{12}_{\vec{q}}\right)^2 + 
    2\left|A^{12}_{\vec{q}}\right|^2 \pm \mathcal{E}^2_{\vec{q}}}\,,
    \label{eq.specs}
\end{eqnarray}
where
\begin{align}
\mathcal{E}^2_{\vec{q}} =
\left[\left(\left(A^{11}_{\vec{q}}\right)^2 - \left(A^{22}_{\vec{q}}\right)^2\right)^2 - 4\left(B^{12}_{\vec{q}}\right)^2 \left(A^{11}_{\vec{q}} - A^{22}_{\vec{q}}\right)^2 \right.\nonumber\\
      \left. + 4\left(A^{11}_{\vec{q}} + A^{22}_{\vec{q}}\right)^2\left|A^{12}_{\vec{q}}\right|^2\right]^{1/2},\nonumber
\end{align}
and the corresponding eigenvectors are denoted by $|\Psi^{\pm}_{\vec{q}}\rangle$, whose explicit forms are provided in Appendix~\ref{app.bosonic-sw}. 
An alternative derivation of the eigenvalues and eigenvectors of the magnon branches, obtained by solving the linearized classical Landau-Lifshitz equations, is presented in Appendix~\ref{app:LL}. Both methods result in the same eigenenergies and eigenvectors.
The spin-wave energies $E^{\pm}_{\vec{q}}\equiv\hbar\omega^{\pm}_{\vec{q}}$ are shown in Fig.~\ref{fig.specs}. 

 In the limit $H=0$, DMI drops out of the spin-wave Hamiltonian Eq.~\eqref{eq.lsw-Ham} and, therefore, does not affect the magnon spectrum, realizing the pure altermagnetic limit. 
In this case, our results reduce to those discussed in Refs.~\cite{Canals2002,Yershov2024,Consoli2025}, where the two magnon branches exhibit $d$-wave-like band-splitting, remaining degenerate along the Brillouin zone diagonals due to the $d$-wave altermagnetic symmetry of the model~\cite{Yershov2024,Consoli2025}. In this limit, magnon branches are physically distinguishable  by the magnon magnetic moment $\vec{\mu}^\pm_{\vec{q}}=-\left.{\mu_0^{-1}}\partial_{\vec{H}}E^{\pm}_{\vec{q}}\right|_{\vec{H}=0}$, which is parallel or antiparallel to the staggered magnetization $(\vec{S}_1-\vec{S}_2)/(2S)$ and has the absolute value $g\mu_B$, see Appendix~\ref{app:H0} for details. 
The importance of taking into account the magnetic moment of magnons when considering altermagnetic splitting has recently been demonstrated for $g$-wave altermagnets~\cite{Kravchuk25}.
For nonzero $H$ and $D$, the branches become fully gapped, as can be seen in Fig.~\subref{fig.specs}{(b)}.

We will now see that the symmetries of the full model discussed in Sec.~\ref{sec.model} are reflected in the magnon spectrum. 
The lower magnon branch exhibits a Goldstone mode at the Brillouin zone center $\rm{\Gamma}$, characterized by a linear dispersion $\cramped{E^{-}_{\vec{q}} \sim |\vec{q}|}$ as $\vec{q} \rightarrow \Gamma$ [see Fig.~\ref{fig.specs}], as a consequence of the spontaneous breaking of the $\rm{U}(1)$ spin-rotation symmetry of the Hamiltonian [Eq.~\eqref{eq.model-Ham}] about the $\vhat{z}$ axis. 
The $C_4$ lattice-rotation about the $\vhat{z}$ axis passing through any $J_1$ bond center, while being a symmetry of the full model as discussed in Sec.~\ref{sec.model}, does not leave the classical ground state ordering [Eq.~\eqref{eq.cgs}] invariant, as it interchanges the magnetic sublattices. 
Nevertheless, the same $C_4$ lattice-rotation combined with $C_2$ spin-rotation about the $\vhat{z}$ axis leaves the ground state ordering as well as the full Hamiltonian invariant. 
Consequently, the linear spin-wave Hamiltonian [Eq.~\eqref{eq.lsw-Ham}] is also invariant under this combined transformation, resulting in the relation $E^\pm_{C_4\vec{q}} = E^\pm_{\vec{q}}$,  i.e., a four-fold rotational symmetry of spectrum in both the branches. 
As seen in Fig.~\eqref{fig.specs}, the spectrum along ${\rm \Gamma X}$ is exactly same as along ${\rm \Gamma Y}$ for both bands.

The upper magnon branch is gapped at the zone center with energy $E^{+}_{\Gamma} = |g\mu_{\textsc{b}}\mu_0H|$. 
The energy gap between the two branches at $\vec{q} = {\rm{M}} \equiv (\pi/a_0,\pi/a_0)$ is set by a combination of DMI and the Zeeman field: $\Delta_{\rm M}=g\mu_B\mu_0|DH|/J$. 
In contrast, at $\vec{q} = {\rm{X}} \equiv (\pi/a_0,0)$ or $\vec{q} = {\rm{Y}} \equiv (0,\pi/a_0)$, the gap is set only by the AM coupling: $\Delta_{\rm X}=4|J_1|S$ 
[see Fig.~\subref{fig.specs}{(b)}]. 
Thus, the gap at X and Y is controlled by altermagnetism, while the gap at M is determined by the combined action of DMI and the Zeeman field. 
The same band gap at X and Y is also a consequence of the four-fold rotation symmetry of the spectrum discussed in the previous paragraph. 
The two magnon bands touching at X/Y for $J_1 =0$ is protected by a combined symmetry. The first operation is the $C_4$ lattice rotation together with a $C_2$ spin rotation described in the previous paragraph, and the second is a mirror reflection about a line along a nearest-neighbor bond. While the former is a symmetry in the presence of $J_1$ bond, the latter is not -- it exchanges the checkerboard and empty plaquettes. Consequently, the combination of these two operations is a symmetry only when $J_1 = 0$. 
Under the first operation, the sublattices are interchanged and $\vec{q}\rightarrow C_4\vec{q}$, and under the subsequent mirror (e.g., about the line along $(\vhat{x}-\vhat{y})/\sqrt{2}$), $C_4\vec{q}\rightarrow (q_x,-q_y)$. For simplicity, we consider points on the Brillouin-zone edge $q_x = \pi/a_0$, where the spin-wave Hamiltonian [Eq.~\eqref{eq.lsw-Ham}] simplifies as all anomalous terms vanish, $B_{(\pi/a_0, \,q_y)}^{\alpha\beta} = 0$. The corresponding magnon spectra are $E^\pm_{(\pi/a_0,\, q_y)}= E_0\left(1\pm \frac{D}{J}\sin(\frac{q_ya_0}{2})\sin\xi\right)$. Under the combined symmetry operation, the Holstein-Primakoff bosons transform as $\l{a}_{(\pi/a_0,\,q_y),\alpha}\rightarrow \l{a}_{(\pi/a_0,-q_y),\alpha'}$ where $\alpha$ and $\alpha'$ denote two different sublattices. This symmetry further enforces $E^\pm_{(\pi/a_0,\, q_y)}= E^\mp_{(\pi/a_0,\,- q_y)}$ which in particular implies at X$=(\pi/a_0,\, 0)$  that $E^\pm_{\rm X}= E^\mp_{\rm X}$, leading to a band touching at X. When $J_1\neq0$, this symmetry is broken and a finite gap opens between the two magnon branches. The same reasoning applies at Y. 
In this work, we will focus more on the coupling parameter regime where the gap at M is already opened by the DMI and Zeeman field, and the AM coupling is tuned from zero to a finite value -- highlighting the role of altermagnetism in opening up the gap between the magnon branches at X and Y. 
The regime of interest in this context would be $\Delta_{\rm X}\ll\Delta_{\rm M}$.

\section{Topology of the magnon bands: Berry curvature and Chern numbers}
\label{sec.Berry-curvature}
\begin{figure}
\includegraphics[width=\columnwidth]{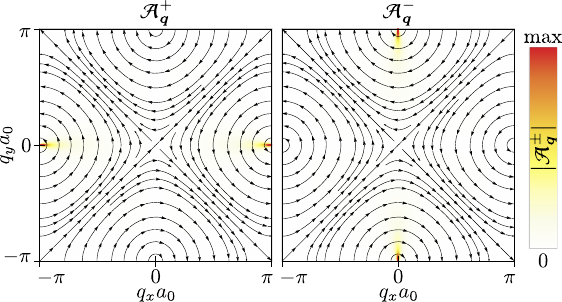}
\caption{Berry connections ${\cal\vec{A}}^\pm_{\vec{q}}$ [Eq.~\eqref{eq.Berry-conn}] of the upper and lower branches for the AM coupling $J_1=-0.04J$, DMI constant $D=0.5J$, and Zeeman field $H=H_s/8$.} 
\label{fig.berry-connection}
\end{figure}

To investigate the topological nature of the gapped magnon branches [i.e., $J_1\neq 0, H\neq 0, D\neq 0$], we begin by calculating the Berry connection associated with each branch. The Berry connection is defined in terms of the corresponding eigenvectors $|\Psi^{\pm}_{\vec{q}}\rangle$ as~\cite{McClarty2022}
\begin{align}
\label{eq.Berry-conn}
{\cal \vec{A}}^\pm_{\vec{q}} = i\langle\Psi_{\vec{q}}^\pm|\mat{\eta}\,\vec{\nabla}_{\vec{q}}|\Psi_{\vec{q}}^{\pm}\rangle =i\left[\begin{array}{c}\langle \Psi_{\vec{q}}^\pm|\mat{\eta}\partial_{q_x}|\Psi_{\vec{q}}^\pm\rangle \\
\langle \Psi_{\vec{q}}^\pm|\mat{\eta}\partial_{q_y}|\Psi_{\vec{q}}^\pm\rangle\end{array}\right],
\end{align}
where 
\begin{equation}\label{eq.eta}
    \mat{\eta} = \left[\begin{array}{cccc}
        1 &  0 & 0 &0\\
         0 &  1 & 0 &0\\
          0 &  0 & -1 &0\\
          0 &  0 & 0 &-1\\
         \end{array}\right]
\end{equation}
is the para-unitary metric that arises due to the bosonic nature of magnons, and the eigenvectors are normalized according to Eq.~\eqref{eq:ev-norm}.
The Berry connection in Eq.~\eqref{eq.Berry-conn} represents a two-component \emph{real-valued} vector field over the two-dimensional wave-vector space. 
Explicit expressions of ${\cal \vec{A}}^\pm_{\vec{q}}$ are given in Appendix~\ref{app.bosonic-sw} [Eqs.~\eqref{eq.B-conn-p} and \eqref{eq.B-conn-m}]. 
Fig.~\ref{fig.berry-connection} shows the Berry connections as vector-field plots in the first Brillouin zone, revealing that the vector-fields in both branches circulate around X and Y points where the connections are \emph{singular} (i.e., divergent and ill-defined).  
These singularities reflect the fact that the magnon wavefunctions $|\Psi_{\vec{q}}^\pm\rangle$ are \emph{not} smooth and single-valued across the entire Brillouin zone. 
This already hints at the possibility that the magnon branches may be topologically non-trivial. 

%%%%%%%%%%%%%%%%%%%%%%%%%%%%%%%%%%
%%%%%%%%%%%%%%%%%%%%%%%%%%%%%%%%%%

\begin{figure*}
\centering
\includegraphics[width=0.95\textwidth]{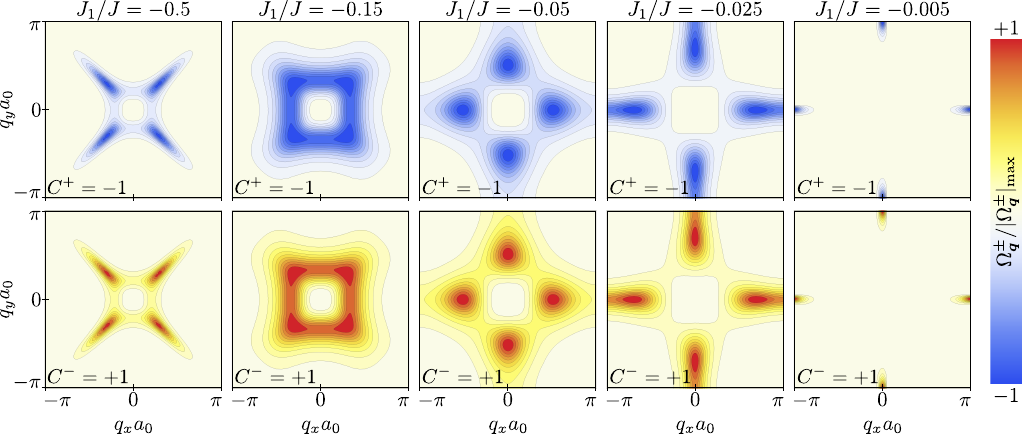}
\caption{Evolution of the Berry curvature [Eq.~\eqref{eq.Berry-curv}] of the upper (top row) and lower (bottom row) magnon branches as the absolute value of the AM coupling $J_1$ decreases from left to right. The other parameters are fixed at the DMI constant $D=0.5J$ and the Zeeman field $H=H_s/8$.} 
\label{fig.berry-curvature}
\end{figure*}

To confirm this, we next calculate the Berry curvature from the Berry connection, defined as 
\begin{align}\label{eq.Berry-curv}
\Omega_{\vec{q}}^\pm=\left(\vec{\nabla}_{\vec{q}}\times\mathcal{\vec{A}}^\pm_{\vec{q}}\right)\cdot\vhat{z}
\end{align}
and show it across the first Brillouin zone for the two branches in Fig.~\ref{fig.berry-curvature} for different values of the AM coupling. 
The $C_4$ lattice rotation combined with a $C_2$ spin-rotation symmetry about the $\vhat{z}$ axis of the Hamiltonian [Eq.~\eqref{eq.model-Ham}] is reflected in the Berry curvature as $\Omega^\pm_{C_4\vec{q}}=\Omega^\pm_{\vec{q}}$, as evident in Fig.~\ref{fig.berry-curvature}.
Applying this symmetry twice successively results in the even-under-inversion property of the Berry curvature, i.e., $\Omega^\pm_{-\vec{q}}=\Omega^\pm_{\vec{q}}$. 
Finally, the topology of the magnon bands in two-dimensional systems is characterized by the (first) Chern number defined in terms of the Berry curvature as 
\begin{equation}\label{eq.Chern}
   C^{\pm} = \frac{1}{2\pi}\int_{\rm FBZ}\Omega_{\vec{q}}^\pm\,\, dq_x\, dq_y,  
\end{equation}
where the integration is performed over the first Brillouin zone (FBZ). 
Evaluating this expression for the two gapped bands yields $|C^{\pm}| = 1$~\footnote{The integral in Eq.~\eqref{eq.Chern} is performed numerically using the analytical form of the Berry curvature for a wide range of parameter sets.} with $C^+=-C^-$, thereby clearly demonstrating that the magnon bands in our model are topologically \emph{non-trivial}. 

We now argue, based on symmetry considerations, that the magnons in the pure altermagnetic limit (i.e., $H = D = 0$) are topologically trivial.
While the Hamiltonian in this limit respects time-reversal symmetry ${\cal T}$, the altermagnetic N\'eel ground state (e.g., aligned along the $\vhat{x}$ axis) breaks it, as the staggered magnetization is reversed. 
However, combining time-reversal with a $C_2$ spin-rotation about the $\vhat{z}$ axis yields an effective time-reversal symmetry, ${\cal T} C_2$, that leaves both the Hamiltonian and the N\'eel ground state invariant. 
This effective time-reversal symmetry imposes the constraint $\Omega^\pm_{\vec{q}} = -\Omega^\pm_{\vec{-q}}$~\cite{Oliveira23,Cheng2016}. 
Additionally the model possesses inversion symmetry which leads to $\Omega^\pm_{\vec{q}} =\Omega^\pm_{\vec{-q}}$~\cite{Oliveira23,Cheng2016}. 
Together, these two constraints force the Berry curvature to vanish identically throughout the Brillouin zone, implying that the magnon bands are topologically trivial in the altermagnetic limit. 
Upon introducing the DMI and Zeeman field, inversion remains a symmetry, but effective time-reversal no longer is, resulting in non-zero Berry curvature and topological Chern bands, as demonstrated in the previous paragraph. 
In other words, the $d$-wave line degeneracies present along the Brillouin zone diagonals in the altermagnetic limit are lifted by the Zeeman field and DMI, driving a transition from topologically trivial to non-trivial magnon bands.

%%%%%%%%%%%%%%%%%%%%%%%%%%%%%%%%%%
%%%%%%%%%%%%%%%%%%%%%%%%%%%%%%%%%%

We find the following interesting properties of the Berry curvature and Chern number for each magnon branch. 
Under the sign reversal of either $H$ or $D$, the spin-wave matrix $\mat{M}_{\vec{q}}$ (see Appendix~\ref{app.bosonic-sw} for details) is complex conjugated, $\mat{\bar{M}}_{\boldsymbol{q}}$. Under complex conjugation, the eigenvalue equation for the magnon branches, $\mat{\eta}\mat{M}_{\vec{q}}|\Psi^{\pm}_{\vec{q}}\rangle = E^{\pm}_{\vec{q}}|\Psi^{\pm}_{\vec{q}}\rangle$ [Eq.~\eqref{eq.EVP-Q} in Appendix~\ref{app.bosonic-sw}] becomes $\mat{\eta}\mat{\bar{M}}_{\vec{q}}|\bar{\Psi}^{\pm}_{\vec{q}}\rangle = E^{\pm}_{\vec{q}}|\bar{\Psi}^{\pm}_{\vec{q}}\rangle$. 
This shows that flipping the sign of $H$ or $D$ leaves the magnon energy eigenvalues unchanged, while the eigenvectors are complex conjugated, $|\bar{\Psi}^{\pm}_{\vec{q}}\rangle$. 
Consequently, the Berry connection changes sign since $\langle\Psi_{\vec{q}}^\pm|\mat{\eta}\,\vec{\nabla}_{\vec{q}}|\Psi_{\vec{q}}^{\pm}\rangle$ is purely imaginary. 
Hence, the Berry curvature [Eq.~\eqref{eq.Berry-curv}] changes as $\Omega^\pm_{\vec{q}}\to-\Omega^\pm_{\vec{q}}$, resulting in $C^\pm\to-C^\pm$. 
When the AM coupling is reversed, $J_1\to -J_1$, the Berry curvature also changes sign at each wave vector, but unlike the previous cases, its absolute value does not remain unchanged. In this case, there is no simple relation between the old and new eigenvectors. Nevertheless, the change happens in such a way that we still find $C^\pm\to-C^\pm$. 

We now examine how the Berry curvature of these topological bands evolves as the AM coupling changes.  
As shown in Fig.~\ref{fig.berry-curvature}, as $|J_1|$ is gradually reduced, the Berry curvature becomes increasingly localized around X and Y points, and nearly vanishes elsewhere in the Brillouin zone.  
In the limit of small $J_1$ (i.e., $\Delta_{\rm X}\ll\Delta_{\rm M}$), the Berry curvature near X point, with $\vec{q} = (\pi/a_0 +k_x, k_y)$, takes the form  
 \begin{equation}\label{eq.Berry-curv-lim1}
\Omega_{\vec{q}}^+ \approx \sgn\left(J_1\right)\frac{DJa_0^2 \sin^3\xi}{2J_1^2}\left[1 - \frac{3J^2a_0^2\sin^2\xi}{2J^2_1}\Xi^2_{\vec{k}}\right],
 \end{equation}
 for $|\vec{k}|a_0\ll|J_1|/J$. Here $\Xi^2_{\vec{k}}=\sin^2\xi\, k_x^2+\frac{D^2}{J^2}k_y^2$.
 From the four-fold rotational relation of the Berry curvature, $\Omega^\pm_{C_4\vec{q}}=\Omega^\pm_{\vec{q}}$, a similar expression holds near Y point, except that $k_x\to k_y$ and $k_y \to -k_x$. 
 The first term in Eq.~\eqref{eq.Berry-curv-lim1} shows that $\Omega_{\vec{q}}^+$ scales as $1/J_1^2$ as $\vec{q}\rightarrow$ X or Y, and the second term controls how sharply the curvature is localized around these points. 
 The area of localization has the shape of an ellipse whose semiaxes in $\vhat{x}$ and $\vhat{y}$ directions are determined by the Zeeman field  ($a_x\propto1/|\sin\xi|$) and DMI ($a_y\propto J/|D|$), respectively.
 
In the same small $J_1$ limit (i.e., $\Delta_{\rm X}\ll\Delta_{\rm M}$), but farther away from X point, i.e., for $|\vec{k}|a_0\gg|J_1|/J$, we find
  \begin{equation}\label{eq.Berry-curv-lim2}
\Omega_{\vec{q}}^+ \approx \sgn\left(H/H_s\right)\frac{DJ_1}{2J^2a_0\Xi^3_{\vec{k}}},
 \end{equation}
 revealing that $\Omega_{\vec{q}}^+$ falls off as $1/|\vec{k}|^3$ as we move sufficiently away from X point. 
  Together, Eqs.~\eqref{eq.Berry-curv-lim1} and \eqref{eq.Berry-curv-lim2} quantitatively characterize the localization of the Berry curvature around X and Y in the limit of weak AM coupling. 
 As $|J_1|\rightarrow 0$, the Berry curvature becomes sharply localized at the X and Y points, approaching Dirac delta functions.  
 An analogous analysis for the lower magnon branch yields expressions identical to Eqs.~\eqref{eq.Berry-curv-lim1} and \eqref{eq.Berry-curv-lim2} up to a sign, consistent with $\Omega_{\vec{q}}^-$ having the opposite sign to $\Omega_{\vec{q}}^+$ across the Brillouin zone as shown in Fig.~\ref{fig.berry-curvature}. 
 An interesting observation is that, in each branch, since the Berry curvature has the same sign at both X and Y points and is nearly zero elsewhere in the limit of weak AM coupling, one may immediately conclude that the Chern number is nonzero (i.e., topological band) without having to perform detailed calculations. 
 Based on Eqs.~\eqref{eq.Berry-curv-lim1} and \eqref{eq.Berry-curv-lim2}, we may obtain the following Berry curvature approximation useful for practical means 
 \begin{equation}\label{eq:Berry-aprox}
     \Omega^+_{\vec{q}}\approx\varpi_{\vec{k}}=\sgn\left(H/H_s\right)\frac{DJ_1}{2J^2a_0}\begin{cases}k_0^{-3}, & \Xi_{\vec{k}}<k_0\\
     \Xi^{-3}_{\vec{k}}, &\Xi_{\vec{k}}>k_0,
     \end{cases}
 \end{equation}
 where constant $k_0=\frac32|J_1/(J\sin\xi)|a_0^{-1}$ is determined from the normalization condition $\frac{1}{2\pi}\iint_{-\infty}^{\infty}|\varpi_{\vec{k}}|\mathrm{d}k_x\mathrm{d}k_y=\frac12$ which guaranties $|C^+|=1$~\footnote{Since the Berry curvature $\varpi_{\vec{k}}$ is symmetrically distributed about X and Y points, contribution to the Chern number from around each point is one-half.}. 
 In other words, the approximated Berry curvature Eq.~\eqref{eq:Berry-aprox} is a constant within the ellipse $\Xi_{\vec{k}}^2=k^2_0$ and coincide with the asymptotic Eq.~\eqref{eq.Berry-curv-lim2} outside it.

Although the Berry curvature appears nearly equal and opposite for the two branches near the center of the Brillouin zone in Fig.~\ref{fig.berry-curvature}, the asymptotic behavior in the vicinity of the zone center ($\Gamma$ point) differs between the branches: 
\begin{align}\label{eq:BCp-Gamma}
    &\Omega_{\vec{q}}^+\approx\frac{\sgn\left(H/H_s\right)J_1D\left(3-\sin^2\xi\right)}{64J^2\sin^2\xi}a_0^4|\vec{q}|^2+O(|\vec{q}|^4),\\
    \label{eq:BCm-Gamma}&\Omega_{\vec{q}}^-\approx-\frac{J_1D\cot\xi}{32J^2\sqrt{1-J_1/J}}a_0^3|\vec{q}|(1+\sin^22\chi)+O(|\vec{q}|^3),
\end{align}
where the angle $\chi$ determines the orientation of the wave vector $\vec{q}$ such that $\tan\chi=q_y/q_x$. 
These estimates are valid in the limit $a_0|\vec{q}|\ll |\sin\xi|$. 
The asymptotics in Eqs.~\eqref{eq:Berry-aprox}, \eqref{eq:BCp-Gamma}, and \eqref{eq:BCm-Gamma} will be useful in the following section where we analytically estimate certain measurable quantities.

\section{Thermal Hall conductivity}
As a physical consequence of the nontrivial topology of the magnon bands, we consider the magnon thermal Hall effect, i.e., the effect of heat transfer by magnons in the direction perpendicular to the applied temperature gradient~\cite{Katsura2010,Matsumoto2014,Onose2010}. 
This phenomenon has been studied extensively in both ferromagnetic~\cite{Katsura2010,Onose2010,Matsumoto11,Ideue2012,Chisnell2015,Hirschberger2015,Murakami17,Mook2014a,Mook2014b,Cao2015,Owerre2016b,Seshadri2018,Malki2019,Pires2021,Zhuo2021} and antiferromagnetic~\cite{Oliveira23,Bhowmick2020,Pires2020,Owerre2016c,Owerre2017,Owerre2017_kag,Laurell2017,Laurell2018,Neumann2022,Chen2023,Mook19,Kawano19} systems, while its realization in an altermagnetic system has only recently begun to attract attention~\cite{Hoyer25}. 
To compute the thermal Hall conductivity, we use the formula~\cite{Katsura2010,Matsumoto2014,Matsumoto11}
\begin{equation}\label{eq:kappaxy}
    \kappa_{xy}=-\frac{k_{\textsc{b}}^2T}{\hbar}\sum\limits_{\nu=\pm}\int_{\mathrm{FBZ}}\frac{dq_x\, dq_y}{(2\pi)^2}c_2(\varrho^\nu_{\vec{q}})\Omega^\nu_{\vec{q}},
\end{equation}
where $k_{\textsc{b}}$ is the Boltzmann constant, $c_2(x)\equiv\int_0^x[\ln(1+t^{-1})]^2\mathrm{d}t$, and $\varrho^\nu_{\vec{q}}=[e^{E^\nu_{\vec{q}}/(k_\textsc{b}T)}-1]^{-1}$ is the Bose distribution function with magnon energies $E^\pm_{\vec{q}}$.

\begin{figure}
    \centering
    \includegraphics[width=\columnwidth]{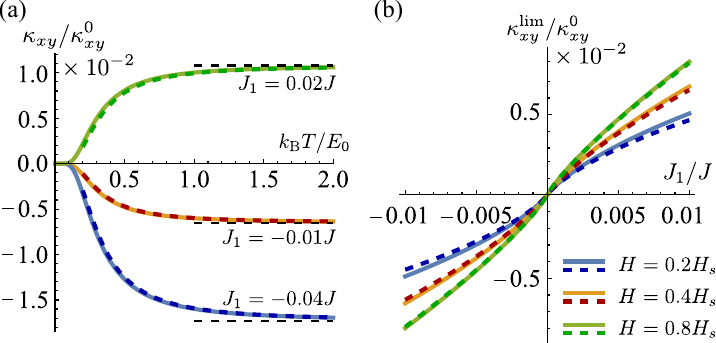}
    \caption{(a) The temperature dependence of the thermal Hall conductivity $\kappa_{xy}$ [Eq.~\eqref{eq:kappaxy}] obtained for the DMI constant $D=0.5J$, Zeeman field $H=0.4H_s$, and three different values of the AM coupling $J_1$. The exact value computed from Eq.~\eqref{eq:kappaxy} and the high-temperature estimation Eq.~\eqref{eq:kappaxy-approx} are shown by solid and thick dashed lines, respectively. The thin horizontal dashed lines show the infinite-temperature asymptotic values $\kappa^{\mathrm{lim}}_{xy}$ determined by Eq.~\eqref{eq:kxy-lim}. (b) Dependence of the infinite-temperature asymptotic value $\kappa^{\mathrm{lim}}_{xy}$ on the AM coupling $J_1$. The DMI constant is taken to be $D = 0.5 J$. Solid and dashed lines correspond to the exact values Eq.~\eqref{eq:kxy-lim} and approximation Eq.~\eqref{eq:kxy-lim-approx}, respectively.}
    \label{fig:kxy-VS-T}
\end{figure}

The dependence of $\kappa_{xy}$ on temperature is shown in Fig.~\subref{fig:kxy-VS-T}{(a)}.
In the low-temperature limit, $k_{\textsc{b}}T\ll|g\mu_{\textsc{b}}\mu_0H|$, the contribution of the magnons from the upper branch $E^+_{\vec{q}}$ can be neglected. The main contribution to $\kappa_{xy}$ comes from the low-energy magnons from the lower branch with energies $E^-_{\vec{q}}~\approx~\frac{1}{2} E_0|\vec{q}|a_0\cos\xi\sqrt{1-J_1/J}$. Using Eq.~\eqref{eq:BCm-Gamma}, we estimate Eq.~\eqref{eq:kappaxy} as
\begin{equation}\label{eq:kappaxy-smlT}
    \kappa_{xy}\approx\kappa^0_{xy}\left(\frac{k_{\textsc{b}}T}{E_0}\right)^4\frac{J_1D}{J^2(1-J_1/J)^2}\frac{c}{\sin\xi\cos^2\xi},
\end{equation}
where $\kappa^0_{xy}=k_{\textsc{b}}E_0/\hbar$, and $c=\frac{15}{2\pi}\zeta(5)\approx2.475$ is a numerical constant with $\zeta(x)$ being the Riemann zeta function. 
Thus, the thermal Hall conductivity at low temperatures scales with temperature as $\kappa_{xy}\propto T^4$. 
Note that in deriving Eq.~\eqref{eq:kappaxy-smlT}, we did not use any assumptions about the smallness of $J_1$ or $D$. 
The power-law temperature dependence of $\kappa_{xy}$ is a consequence of the absence of a gap in the magnon spectrum $E^-_{\vec{q}}$. 
This is in contrast to the exponentially suppressed low-temperature thermal Hall conductivity previously predicted for ferromagnetic Lu$_2$V$_2$O$_7$~\cite{Onose2010}, whose magnon spectrum was assumed to be gapped. On the other hand, our result is similar to the  $T^{7/2}$ scaling previously obtained for the pyrochlore ferromagnet~\cite{Ideue2012,Kawano19} in the gapless limit, i.e., for the vanishing Zeeman field. 

In the infinite-temperature limit, $\kappa_{xy}$ saturates to the asymptotic value~\cite{Mook2014a}
\begin{equation}\label{eq:kxy-lim}
\kappa^{\mathrm{lim}}_{xy}\equiv\lim\limits_{T\to\infty}\kappa_{xy}=\frac{k_{\textsc{b}}}{\hbar}\sum\limits_{\nu=\pm}\int_{\mathrm{FBZ}}\frac{dq_x\,dq_y}{(2\pi)^2}E^\nu_{\vec{q}}\Omega^\nu_{\vec{q}},
\end{equation}
shown by the horizontal dashed lines in Fig.~\subref{fig:kxy-VS-T}{(a)}. 
At sufficiently high temperatures, when magnons in both the branches are thermally excited, one can analytically estimate $\kappa_{xy}$ in the regime $\Delta_{\rm X}\ll\Delta_{\rm M}\ll J$. 
In this regime, the Berry curvature is localized in the vicinity of X and Y points and $\Omega^+_{\vec{q}}=-\Omega^-_{\vec{q}}$, as discussed in Sec.~\ref{sec.Berry-curvature}. 
This yields
\begin{equation}\label{eq:kappaxy-approx}
    \kappa_{xy}\approx\kappa^{\mathrm{lim}}_{xy}\left[\frac{\beta'/2}{\sinh(\beta'/2)}\right]^2,\qquad\beta'=\frac{E_0}{k_BT}.
\end{equation}
The estimation in Eq.~\eqref{eq:kappaxy-approx} is shown in Fig.~\subref{fig:kxy-VS-T}{(a)} by thick dashed lines. 
In the considered regime, we can use the Berry curvature approximation in Eq.~\eqref{eq:Berry-aprox} to estimate the infinite-temperature limit
\begin{equation}\label{eq:kxy-lim-approx}
    \kappa^{\mathrm{lim}}_{xy}\approx \kappa^0_{xy}\frac{J_1/J}{2\pi}\varsigma\left[\frac13+\ln\left|\frac{4\pi D\sin^2\xi}{3J_1\left(|\sin\xi|+|D|/J\right)}\right|\right],
\end{equation}
where $\varsigma=\sgn\left(\frac{H}{H_s}\frac{D}{J}\right)$. Fig.~\subref{fig:kxy-VS-T}{(b)} demonstrates a very good agreement between the exact value Eq.~\eqref{eq:kxy-lim} (solid lines) and the estimation Eq.~\eqref{eq:kxy-lim-approx} (dashed lines) of $\kappa^{\mathrm{lim}}_{xy}$ in the regime $\Delta_{\rm X}\ll\Delta_{\rm M}\ll J$. 

\begin{figure}
    \centering
    \includegraphics[width=\columnwidth]{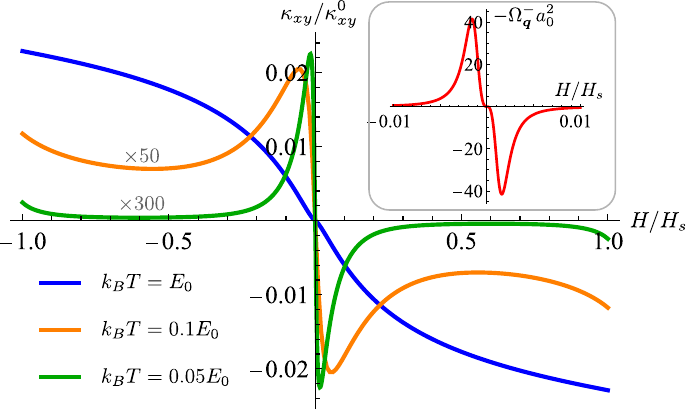}
    \caption{The Zeeman field dependence of the thermal Hall conductivity [Eq.~\eqref{eq:kappaxy}] for the DMI constant $D = 0.5J$, AM coupling $J_1 = -0.04 J$, and different temperatures.  The values for the orange and green curves are magnified 50 and 300 times, respectively. The inset shows the field dependence of the Berry curvature of the lower magnon branch in the vicinity of $\Gamma$ point: $(q_x,q_y)=(0.1,0)/a_0$.} 
    \label{fig:kappa-VS-H}
\end{figure}
The Zeeman field dependence of $\kappa_{xy}$ is shown in Fig.~\ref{fig:kappa-VS-H}. As the field is reversed, $\kappa_{xy}$ changes sign and, notably, exhibits a sharp jump across zero at low temperatures. 
This non-monotonic behavior of $\kappa_{xy}$ for low fields at low temperatures follows from the field dependence of the Berry  curvature $\Omega^-_{\vec{q}}$ of the lower magnon branch near $\Gamma$ point, see the inset in Fig.~\ref{fig:kappa-VS-H}. 
Note that the magnons in the lower branch from the vicinity of $\Gamma$ point mainly contribute to $\kappa_{xy}$ at low temperatures. 

\section{Edge states}

%%%%%%%%%%%%%%%%%%%%%%%%%%%%%%%%%%
\begin{figure}
\centering
\includegraphics[width=\columnwidth]{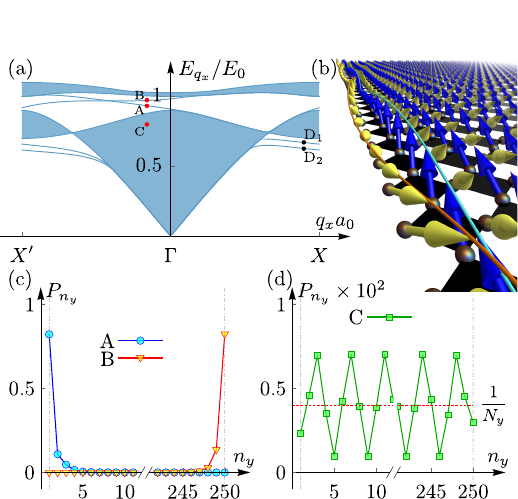}
\caption{(a) Magnon spectrum of our model on a strip [by diagonalizing $\mat{H}_{q_x}^{\rm str}$ in Eq.~\eqref{eq:Psi-edge}] with $N_y=250$ shells in the $\vhat{y}$ direction obtained for the DMI constant $D = 0.4J$, AM coupling $J_1 = 0.1 J$, and Zeeman field $H = H_s/2$. Panel (b) shows a snapshot of the reconstructed dynamics of the magnetic moments of the edge state, corresponding to point A (for details, see Appendix~\ref{app:LL-edge} and SM video \cite{SM}). Panels (c) and (d) show the distribution of wavefunction intensity in Eq.~\eqref{eq:intens} across the shells for the points A,B, and C, respectively.} 
\label{fig.edge}
\end{figure}
%%%%%%%%%%%%%%%%%%%%%%%%%%%%%%%%%%
Another physical consequence of the nontrivial topology of the magnon bands is the existence of edge states in finite geometries, which reflects the bulk-boundary correspondence principle~\cite{McClarty2022,Zhang2013}. 
To investigate these edge modes in our altermagnetic model, we consider the system on a strip geometry: the system is infinite along the $\vhat{x}$ direction and finite along the $\vhat{y}$ direction comprising $N_y$ shells, 
for details see Appendix~\ref{app:LL-edge}. 
This setup preserves discrete translational symmetry along the $\vhat{x}$ direction, but not along the $\vhat{y}$ direction. 
Applying the Fourier transform for the $\vhat{x}$ direction, we introduce $q_x$, and we leave the real-space representation for the $\vhat{y}$ direction. 
Thus, the linear spin-wave Hamiltonian for the strip is represented by $4N_y\times 4N_y$ matrix $\vec{H}^{\mathrm{str}}_{q_x}$ [see Appendix~\ref{app:LL-edge} for details]. 
Fig.~\subref{fig.edge}{(a)} shows the numerically obtained spectrum $E_{q_x}$ computed for the strip geometry with a large number of shells $N_y$ in the $\vhat{y}$ direction. 
The two prominent continua correspond to the two bulk magnon bands previously discussed in Sec.~\ref{sec.swt-magnon-bands}. 
Importantly, additional \emph{in-gap} states appear between these continua and connect the bulk bands. These in-gap states correspond to the topologically protected excitations localized at the edges of the system~\cite{McClarty2022,Mook2016}.

To confirm the localized character of these states at the edges, we examine the distribution of their wavefunction intensity 
\begin{equation}\label{eq:intens}
    P_{n_y}=S\sum_{\nu=1,2}\left(|\psi_{q_x,n_y,\nu}|^2+|\psi_{-q_x,n_y,\nu}|^2\right)
\end{equation}
across the strip. Here, functions $\psi_{q_x,n_y,\nu}$ and $\psi_{-q_x,n_y,\nu}$ encode the amplitude and polarization of the magnetic moments precession, see Appendix~\ref{app:LL-edge} for details. 
To compare the intensity distributions of the wavefunctions $P_{n_y}$ over the strip for different points (A, B, C) we utilize the normalization rule $\sum_{n_y=1}^{N_y}P_{n_y}=1$ for all cases. 
As shown in Figs.~\subref{fig.edge}{(c)} and \subref{fig.edge}{(d)}, representative in-gap states [points A and B in Fig.~\subref{fig.edge}{(a)}] are localized at the different edges of the strip, while a bulk state [e.g., point C in Fig.~\subref{fig.edge}{(a)}] exhibits almost uniform distribution across the full width, with the average intensity $\overline{P}_{n_y}=1/N_y$. Additionally, using Eq.~\eqref{eq:m-reconstr} we can reconstruct the real-space dynamics of the magnetic moments driven by Landau-Lifshitz equations for the edge and bulk states marked in Fig.~\subref{fig.edge}{(a)} by points A and C, respectively (see SM videos~\cite{SM}). 

The direction of the magnon group velocity $\hbar^{-1}\partial_{q_x}E_{q_x}$  reveals that the topologically protected edge states on the opposite edges of the strip propagate in opposite directions [see Fig.~\subref{fig.edge}{(a)}]. This unidirectional (chiral) propagation of the edge modes at each edge reflects the non-zero topological Chern numbers $C^{\pm}=\mp 1$ associated with the bulk magnon bands~\cite{McClarty2022,Zhang2013,Malki2019}.

One should note that the additional edge states below the lower bulk band, e.g., marked with black dots D$_1$ and D$_2$ in  Fig.~\subref{fig.edge}{(a)} arise from the effective edge potential due to the lack of neighbors for the spins at the edges. In contrast to the edge states A and B, the edge states D$_1$ and D$_2$ do not connect the two bulk bands with different Chern numbers, and therefore they are \emph{not} topologically protected~\cite{Mook2014b}.

\section{Conclusions}
We have explored topological magnons in a minimal altermagnetic spin model -- the checkerboard lattice spin model in the presence of a Zeeman field and DMI. 
These interactions stabilize a long-range ordered coplanar spin-flopped state, with spins canted out of the lattice plane towards the direction of the Zeeman field. 
Using linear spin-wave theory, we have analyzed the magnon excitations about this ordered state and found that nonzero AM coupling, Zeeman field, and DMI open a gap between the two magnon bands, rendering them topologically non-trivial with Chern numbers $|C^{\pm}| = 1$, and $C^+=-C^-$. 
Reversing the sign of any of the three couplings -- AM coupling, Zeeman field, or DMI flips the sign of the Chern numbers. 
In the weak AM coupling limit, we find that the Berry curvature becomes sharply localized around the X and Y points in the Brillouin zone, forming elliptical profiles whose semi-axes are individually controlled by the Zeeman field and DMI, respectively.

As a direct consequence of the topological character of magnons, we find a nonzero thermal Hall response under an applied temperature gradient.  
At low temperatures, thermal Hall conductivity $\kappa_{xy}$ scales with temperature as $\kappa_{xy}\propto T^4$.  
Moreover, $\kappa_{xy}$ changes sign upon reversing the direction of the Zeeman field, displaying a sharp jump across zero field at low temperatures.  
We have also demonstrated the bulk-boundary correspondence by considering our model on a finite strip geometry, which reveals topologically protected unidirectional (chiral) spin-wave modes localized at the edges.

We have focused here on a minimal model where the next-nearest-neighbor couplings act along a single direction per sublattice, i.e., a nonzero coupling $J_1$ within the checkerboard plaquettes and none in the empty ones. One can, however, also include an additional coupling $J_2$ along the diagonals of the empty plaquettes. As long as $J_2\neq J_1$, the system continues to exhibit altermagnetism, i.e., in the limit of $H = D=0$, the two sublattices can still be mapped to one another by $C_4$ lattice-rotation about the $\vhat{z}$ axis passing through either a $J_1$ or $J_2$ bond center combined with time-reversal flipping the spins. Essentially, the model has same symmetries for $J_2 = 0$ case and the case of nonzero $J_2 \neq J_1 $. We find that the model still hosts topological magnons with Chern numbers $|C^\pm| = 1$, nonzero thermal Hall response, and chiral edge modes in a finite strip geometry. The gap between the magnon branches at X or Y now depends on the difference between the two couplings ($|J_1 - J_2|$), which quantifies the strength of altermagnetism. While the thermal Hall conductivity may change quantitatively from the $J_2 =0$ case, all qualitative physics of topology and transport remain unchanged.

Typically, in spin models, topological magnon excitations are driven either by DMI or a combination of DMI and Zeeman field~\cite{McClarty2022}. However, this work demonstrates that, in addition to DMI and the Zeeman field, altermagnetism (i.e., via the AM coupling) plays an equally crucial role in generating the non-trivial topology of the magnon bands. 
Beyond its fundamental interest, our study may offer insights for spintronic applications, where the interplay between topologically protected magnons and altermagnetism is particularly important~\cite{Chumak2015,Wang2018,Smejkal2022b}.

Our analysis has been based on linear spin-wave theory, where magnons are treated as non-interacting quasiparticles. 
This approximation is particularly well-justified in the semiclassical large spin-length ($S$) limit at low temperatures, where the density of excited magnons remains low. 
However, in the small-$S$ regime, particularly about non-collinear magnetic orderings, magnon-magnon interactions may become significant and cannot always be safely ignored~\cite{Zhitomirsky2013}. In such cases, it is important to investigate how interactions may modify the topological properties of magnons~\cite{McClarty2022,Chen2023,Chernyshev2016,McClarty2019} -- a direction we leave for future study.

\begin{acknowledgments}
We thank F. F. Assaad, P. M. C\^onsoli, A. Hickey, O. Janson, T. Sato for fruitful discussions. S.~K., V.~K., and J.~v.~d.~B. acknowledge financial support from the Deutsche Forschungsgemeinschaft (DFG, German Research Foundation) under Germany’s Excellence Strategy through the W\"urzburg-Dresden Cluster of Excellence on Complexity and Topology in Quantum Matter - ct.qmat (EXC 2147, project-ids 390858490 and 392019). J.~v.~d.~B. also acknowledges support from the Deutsche Forschungsgemeinschaft (DFG, German Research Foundation) within the Collaborative Research Center ``Correlated Magnetism: From Frustration to Topology'' (SFB 1143, project-id 247310070). K. Y. acknowledges financial support by the Deutsche Forschungsgemeinschaft (DFG, project-id YE 232/2-1). 
\end{acknowledgments}

\appendix

\section{Quantum linear spin-wave Hamiltonian and its diagonalization}
\label{app.bosonic-sw}
To perform spin-wave analysis of the Hamiltonian in Eq.~\eqref{eq.model-Ham} about the classical ground state given in Eq.~\eqref{eq.cgs}, we introduce two \emph{local} right-handed coordinate frames, $(\vhat{e}_{1x},\vhat{e}_{1y}, \vhat{e}_{1z})$ and $(\vhat{e}_{2x},\vhat{e}_{2y}, \vhat{e}_{2z})$, aligned with the spin-directions on sublattice-1 and sublattice-2, respectively, in the ground state magnetic ordering. 
These frames are given by   
\begin{align}
    \vhat{e}_{1x} &= -\sin\xi\, \vhat{x} + \cos\xi\,\vhat{z},\,
    \vhat{e}_{1y} = -\vhat{y},\,
    \vhat{e}_{1z} = \cos\xi \,\vhat{x} + \sin\xi\,\vhat{z},\nonumber\\
    \vhat{e}_{2x} &= \sin\xi \,\vhat{x} + \cos\xi\,\vhat{z},\,
    \vhat{e}_{2y} = \vhat{y},\,
    \vhat{e}_{2z} = -\cos\xi\, \vhat{x} + \sin\xi\,\vhat{z}.
    \label{eq.loc-frame}
\end{align}
Here, $\vhat{e}_{1z}$ and $\vhat{e}_{2z}$ point along the classical ground state spin-directions on sublattice-1 and sublattice-2, respectively. 
Out of these local directions, we can further define 
\begin{equation}
\vhat{e}_{\alpha{\pm}} = (\vhat{e}_{\alpha x} \pm i\, \vhat{e}_{\alpha y})/\sqrt{2}.
\label{app.local-frame}
\end{equation}
where $\alpha = 1, 2$. 
We now map the spins to bosons using the Holstein-Primakoff transformation~\cite{Holstein1940}, 
\begin{align}
    \vec{S}_{\vec{\rho},\alpha} &= \sqrt{S}\left[\left(1-\frac{n_{\vec{\rho},\alpha}}{2S}\right)^{1/2} a_{\vec{\rho},\alpha}\,\vhat{e}_{\alpha-} + a^{\dagger}_{\vec{\rho},\alpha}\left(1-\frac{n_{\vec{\rho},\alpha}}{2S}\right)^{1/2}\vhat{e}_{\alpha +}\right]\nonumber\\
    &\hspace{0.5 cm} +(S-n_{\vec{\rho},\alpha})\, \vhat{e}_{\alpha z},
    \label{eq.HP}
\end{align}
where $\vec{\rho} = \vec{r}$ for $\alpha =1$ and $\vec{\rho} = \vec{r}'$ for $\alpha =2$. The operator $a_{\vec{\rho},\alpha} (a^{\dagger}_{\vec{\rho},\alpha})$ represents the bosonic annihilation (creation) operator at $\vec{\rho}^{\rm{th}}$ site of sublattice-$\alpha$ and number operator $n_{\vec{\rho},\alpha} \equiv a^{\dagger}_{\vec{\rho},\alpha}a_{\vec{\rho},\alpha}$. 
These operators satisfy the commutation relations
\begin{align}
[a_{\vec{\rho},\alpha}, a^{\dagger}_{\vec{\rho}',\beta}] &=\delta_{\vec{\rho},\vec{\rho}'}\delta_{\alpha,\beta},\,\,
     [a_{\vec{\rho},\alpha}, a_{\vec{\rho}',\beta}] =  [a^\dagger_{\vec{\rho},\alpha}, a^\dagger_{\vec{\rho}',\beta}]=0.
\end{align}
Expanding the square roots in Eq.~\eqref{eq.HP} in powers of $\frac{n_{\vec{\rho},\alpha}}{2S}$, and substituting into the Hamiltonian [Eq.~\eqref{eq.model-Ham}] yields
\begin{equation}
    {\cal H} = E_{\rm cl} + {\cal H}_1 + {\cal H}_2 + \cdots,
    \label{eq.Ham-expan}
\end{equation}
where $E_{\rm cl}\sim O(S^2)$ is the classical ground state energy, and ${\cal H}_1\sim O(S^{3/2})$ is linear in bosonic operators and thus vanishes due to the stability of the classical ground state configuration. 
The leading quantum correction arises from ${\cal H}_2 \sim O(S)$ which is bilinear in bosonic operators. 
${\cal H}_2$ is referred to as the linear spin-wave Hamiltonian. 
After a Fourier transform
\begin{equation}\label{eq.FT}
   a_{\vec{q},\alpha} = \frac{1}{\sqrt{N}}\sum_{\vec{\rho}}e^{-i\vec{q}.\vec{\rho}} a_{\vec{\rho},\alpha}, 
\end{equation}
with $N$ being the total number of spins in one sublattice, this takes the form 
\begin{equation}
     {\cal H}_2 = \sum_{\alpha\beta,\vec{q}}\left[{A}^{\alpha\beta}_{\vec{q}}\h{a}_{\vec{q},\alpha}\l{a}_{\vec{q},\beta} +
          \frac{1}{2} \left({B}^{\alpha\beta}_{\vec{q}}\h{a}_{\vec{q},\alpha}\h{a}_{-\vec{q},\beta} +
          \bar{B}^{\alpha\beta}_{\vec{q}}\l{a}_{-\vec{q},\beta}\l{a}_{\vec{q},\alpha}\right)
          \right].
          \label{eq.HP-2}
\end{equation}
The coefficients $A_{\vec{q}}^{\alpha\beta}$ and $B_{\vec{q}}^{\alpha\beta}$ are given in Eq.~\eqref{eq.swH-coeff} in Sec.~\ref{sec.swt-magnon-bands}. 
This Hamiltonian can be rewritten as 
\begin{align}
    {\cal H}_2
    &= \frac{1}{2}\sum_{\vec{q}}\mat{X}_{\vec{q}}^\dagger \mat{M}_{\vec{q}}\mat{X}_{\vec{q}}- \frac{1}{2}\sum_{\vec{q}}\tr{\mat{A}}_{\vec{q}},
    \label{eq.x-M-x}
\end{align}
where $\mat{X}_{\vec{q}}^\dagger = \left[\h{a}_{\vec{q},1}\,\, \h{a}_{\vec{q},2} \,\,\l{a}_{-\vec{q},1}\,\, \l{a}_{-\vec{q},2} \right]$ and  
\begin{equation}\label{eq.M}
    \mat{M}_{\vec{q}} = \left[\begin{array}{cc}
        \mat{A}_{\vec{q}} &  \mat{B}_{\vec{q}}\\
        \mat{B}^\dagger_{\vec{q}} & \mat{A}^\intercal_{-\vec{q}} \end{array}\right].
\end{equation}
$\mat{A}_{\vec{q}}, \mat{B}_{\vec{q}}$ are $2\times 2$ matrices with entries $A^{\alpha\beta}_{\vec{q}}, B^{\alpha\beta}_{\vec{q}}$ respectively. 
Note that $\mat{M}_{\vec{q}}$ is a Hermitian matrix as ${\cal H}_2$ is Hermitian. 
The additional trace term, $- \frac{1}{2}\sum_{\vec{q}}\tr{\mat{A}}_{\vec{q}}$, arising from the bosonic commutation relations, is a $O(S)$ constant quantum contribution to the classical ground state energy $E_{\rm cl}$.

\subsection{Diagonalization of ${\cal H}_2$}\label{sbsc:diagonal}
The linear spin-wave Hamiltonian can be diagonalized by diagonalizing  $\mat{M}_{\vec{q}}$. 
Since this is a bosonic Hamiltonian, it \emph{cannot} be diagonalized using a unitary matrix. Instead, the diagonalization requires a \emph{para-unitary} transformation $\mat{T}_{\vec{q}}$~\cite{Kowalska1966,Blaizot1986} such that $\mat{X}_{\vec{q}} = \mat{T}_{\vec{q}}\mat{\Gamma}_{\vec{q}}$ where $\mat{\Gamma}_{\vec{q}} = \left[\l{\gamma}_{\vec{q},1}\,\, \l{\gamma}_{\vec{q},2}\,\,\h{\gamma}_{-\vec{q},1}\,\, \h{\gamma}_{-\vec{q},2} \right]^{\intercal}$ contains the transformed bosonic operators and $\mat{T}_{\vec{q}}$ satisfies the para-unitary condition $\mat{T}^\dagger_{\vec{q}}\mat{\eta}\mat{T}_{\vec{q}} = \mat{\eta}$ with $\mat{\eta}$ a diagonal matrix defined in Eq.~\eqref{eq.eta}. 
Following Ref.~\cite{Kowalska1966}, we will find such a transformation in three steps.

\subsubsection{Step-I}
From the matrix elements of $\mat{M}_{\vec{q}}$, we define 
\begin{align}
    Q_1 &= \frac{A^{11}_{\vec{q}}- A^{22}_{\vec{q}}}{2}+ \sqrt{\left(\frac{A^{11}_{\vec{q}}+ A^{22}_{\vec{q}}}{2}\right)^2 - \left(B^{12}_{\vec{q}}\right)^2},\nonumber\\
    Q_2 &= \frac{A^{22}_{\vec{q}}- A^{11}_{\vec{q}}}{2}+ \sqrt{\left(\frac{A^{11}_{\vec{q}}+ A^{22}_{\vec{q}}}{2}\right)^2 - \left(B^{12}_{\vec{q}}\right)^2},
    \label{eq.Q1-Q2}
\end{align}
and consider a transformation 
\begin{align}
    \mat{T}_{I} = \left[\begin{array}{cccc}
        r_1 &  0 & 0& r_2\\
         0 &  r_3 & r_4 & 0\\
          0 &  r_5 & r_6& 0\\
           r_7 &  0 & 0& r_8
        \end{array}\right],
\end{align}
where $r_1 = r_6=\sqrt{\frac{A^{11}_{\vec{q}} + Q_2}{Q_1 + Q_2}}$, $r_2 = \bar{r}_5= \sqrt{\frac{A^{11}_{\vec{q}} - Q_1}{Q_1 + Q_2}}\,\frac{A^{12}_{\vec{q}}}{\left|A^{12}_{\vec{q}}\right|}$, $r_3 = \bar{r}_8=-r_1\,\frac{A^{21}_{\vec{q}}}{\left|A^{12}_{\vec{q}}\right|}$, $r_4 = r_7=- |r_2|$. 
This transformation satisfies $\mat{T}_I^{\dagger}\mat{\eta} \mat{T}_I = \mat{\eta}$ and transforms $\mat{M}_{\vec{q}}$ to 
\begin{align}
    \mat{M}^I_{\vec{q}} = \mat{T}_I^{\dagger}\mat{M}_{\vec{q}} \mat{T}_I=  \left[\begin{array}{cccc}
        Q_1 &  P & -\bar{U}& 0\\
        P&  Q_2 & 0& -\bar{U}\\
           -U &  0 & Q_1&P \\
           0 & -U  & P & Q_2
        \end{array}\right],
\end{align}
with $P = -\left(r_1^2+|r_2|^2\right)\left|A^{12}_{\vec{q}}\right|$ and $U = 2r_1|r_2|\,A^{21}_{\vec{q}}$. 

\subsubsection{Step-II}

We next define
\begin{align}
    W_1 &= \frac{Q_1+ Q_2}{2}+ \sqrt{\left(\frac{Q_1-Q_2}{2}\right)^2 + P^2},\nonumber\\
    W_2 &= \frac{Q_1+ Q_2}{2}- \sqrt{\left(\frac{Q_1-Q_2}{2}\right)^2 +P^2},
    \label{eq.W1-W2}
\end{align}
and consider the transformation 
\begin{align}
    \mat{T}_{II} = \left[\begin{array}{cccc}
        s_1 &  s_2 & 0& 0\\
         s_3 &  s_4 & 0 & 0\\
          0 &  0 & s_5& s_6\\
           0 &  0 & s_7& s_8
        \end{array}\right],
\end{align}
where $s_1 =s_4 =s_5= s_8= \sqrt{\frac{Q_1 - W_2}{W_1 - W_2}}$, $s_2=s_6 =-s_3= -s_7 = \sqrt{\frac{W_1 - Q_1}{W_1 - W_2}}$. 
This also satisfies $\mat{T}_{II}^{\dagger}\mat{\eta} \mat{T}_{II} = \mat{\eta}$ and we have 
\begin{align}
 \mat{M}_{\vec{q}}^{II} = \mat{T}_{II}^{\dagger}\mat{M}^I_{\vec{q}} \mat{T}_{II} = \left[\begin{array}{cccc}
        W_1 &  0 &-\bar{U} & 0\\
        0&  W_2 & 0&- \bar{U}\\
           -U &  0 & W_1&0 \\
           0 &- U  & 0 & W_2
        \end{array}\right].
\end{align}

\subsubsection{Step-III}
In this final step, we define 
\begin{align}
    E_+ &= \sqrt{W_1^2 - |U|^2},\nonumber\\
    E_- &= \sqrt{W_2^2 - |U|^2},
    \label{eq.E1-E2}
\end{align}
and take the transformation 
\begin{align}
    \mat{T}_{III} = \left[\begin{array}{cccc}
        t_1 & 0& t_2 & 0\\
         0 &  t_3 & 0 & t_4\\
        t_5 & 0& t_6&0\\
          0& t_7 &  0 & t_8
        \end{array}\right],
\end{align}
where $t_1 = \sqrt{\frac{W_1 + E_+}{2E_+}}$, $t_2 = \sqrt{\frac{W_1 - E_+}{2E_+}}$, $t_3 = \sqrt{\frac{W_2 + E_-}{2E_-}}$, $t_4 = \sqrt{\frac{W_2 - E_-}{2E_-}}$, $t_5 = t_2 \frac{A_{\vec{q}}^{21}}{\left|A_{\vec{q}}^{21}\right|} $, $t_6 =  t_1\frac{A_{\vec{q}}^{21}}{\left|A_{\vec{q}}^{21}\right|}$, $t_7 =  t_4\frac{A_{\vec{q}}^{21}}{\left|A_{\vec{q}}^{21}\right|} $, and $t_8 = t_3\frac{A_{\vec{q}}^{21}}{\left|A_{\vec{q}}^{21}\right|} $. 
This also satisfies the para-unitary condition $\mat{T}_{III}^{\dagger}\mat{\eta} \mat{T}_{III} = \mat{\eta}$, yielding 
\begin{align}
 \mat{M}_{\vec{q}}^{D} = \mat{T}_{III}^{\dagger}\mat{M}^{II}_{\vec{q}} \mat{T}_{III} = \left[\begin{array}{cccc}
        E_+ &  0 &0 & 0\\
        0&  E_- & 0& 0\\
           0 &  0 & E_+&0 \\
           0 & 0  & 0 & E_-
        \end{array}\right].
\end{align} 
Therefore, the complete transformation that diagonalizes the spin-wave Hamiltonian is $\mat{T}_{\vec{q}} = \mat{T}_{I}\mat{T}_{II}\mat{T}_{III}$ and it is easy to see that $\mat{T}_{\vec{q}}^{\dagger}\mat{\eta} \mat{T}_{\vec{q}} = \mat{\eta}$. 
The spin-wave Hamiltonian [Eq.~\eqref{eq.x-M-x}], apart from the constant trace term, under the above three successive transformations becomes 
\begin{align}
&{\cal H}_2
    = \frac{1}{2}\sum_{\vec{q}}\mat{\Gamma}_{\vec{q}}^\dagger\mat{T}_{\vec{q}}^\dagger \mat{M}_{\vec{q}}\mat{T}_{\vec{q}}\mat{\Gamma}_{\vec{q}}\nonumber\\
    &=\frac{1}{2}\sum_{\vec{q}}\left[\h{\gamma}_{\vec{q},1}\,\, \h{\gamma}_{\vec{q},2}\,\,\l{\gamma}_{-\vec{q},1}\,\, \l{\gamma}_{-\vec{q},2} \right] \left[\begin{array}{cccc}
        E_+ &  0 &0 & 0\\
        0&  E_- & 0& 0\\
           0 &  0 & E_+&0 \\
           0 & 0  & 0 & E_-
        \end{array}\right] \left[\begin{array}{c}
        \l{\gamma}_{\vec{q},1}\\
        \l{\gamma}_{\vec{q},2}\\
          \h{\gamma}_{-\vec{q},1}  \\
          \h{\gamma}_{-\vec{q},2} 
        \end{array}\right]\nonumber\\
    &= \sum_{\vec{q}} \left[E_+ \left(\h{\gamma}_{\vec{q},1}\l{\gamma}_{\vec{q},1} +\frac{1}{2}\right)+ E_-\left( \h{\gamma}_{\vec{q},2}\l{\gamma}_{\vec{q},2} + \frac{1}{2}\right)\right].
\end{align}
We have made use of the bosonic commutation relation in the final step.  
The magnon modes are thus described by $\l{\gamma}_{\vec{q},\alpha}$ bosons and the magnon energies are given by $E_+$ and $E_-$. 
In the notations of the main text, $\hbar\omega_{\vec{q}}^+ \equiv E_+$ and $\hbar\omega_{\vec{q}}^- \equiv E_-$, with their explicit expressions given in Eq.~\eqref{eq.specs}. 

Diagonalizing a spin-wave Hamiltonian via a para-unitary transformation can be recast as a standard eigenvalue problem. From the para-unitary condition, we have $\mat{T}_{\vec{q}}^{\dagger}  = \mat{\eta}\mat{T}^{-1}_{\vec{q}}\mat{\eta}$, and thus the transformation $\mat{T}^{\dagger}_{\vec{q}}\mat{M}_{\vec{q}}\mat{T}_{\vec{q}}  = \mat{M}_{\vec{q}}^{D}
$ leads to the eigenvalue equation 
\begin{align} \label{eq.EVP-Q}\left(\mat{\eta}\mat{M}_{\vec{q}}\right)\mat{T}_{\vec{q}}  = \mat{T}_{\vec{q}}\left(\mat{\eta}\mat{M}_{\vec{q}}^{D}\right).
\end{align}
 Therefore, the columns of $\mat{T}_{\vec{q}}$ are the eigenvectors of $\mat{\eta}\mat{M}_{\vec{q}}$, known as the Bogoliubov matrix~\cite{Blaizot1986}, with eigenvalues $(E_+, E_-, -E_+, -E_-)$. We denote these four eigenvectors as $|\Psi_{\vec{q}}^+\rangle, |\Psi_{\vec{q}}^-\rangle $, $|\Phi_{\vec{q}}^+\rangle$, and $|\Phi_{\vec{q}}^-\rangle$. Their orthonormality relations follow from the para-unitary condition satisfied by $\mat{T}_{\vec{q}}$, 
\begin{align}\label{eq:ev-norm}
 \langle\Psi^{\mu}_{\vec{q}}|\mat{\eta}|\Psi^\nu_{\vec{q}}\rangle &= +\,\delta_{\mu\nu},\,\, 
\langle\Phi^\mu_{\vec{q}}|\mat{\eta}|\Phi^\nu_{\vec{q}}\rangle = -\,\delta_{\mu\nu},\,\,
\langle\Psi^\mu_{\vec{q}}|\mat{\eta}|\Phi^\nu_{\vec{q}}\rangle = 0,
 \end{align}
 where $\mu, \nu = \pm$.

\subsection{Berry connection}\label{sbsc:BC}
The Berry connection associated with a magnon branch is defined in Eq.~\eqref{eq.Berry-conn} in terms of its corresponding eigenvector. 
The eigenvector for the upper branch,  $|\Psi_{\vec{q}}^+\rangle$, is given by the first column of $\mat{T}_{\vec{q}}$,  
\begin{align}\label{eq:Psi-p}
|\Psi_{\vec{q}}^+\rangle =  \left[\begin{array}{c}
        r_1s_1t_1 - r_2s_2t_5 \\
       -r_3s_2t_1 + r_4s_1t_5\\
          -\bar{r}_2s_2t_1+r_1s_1t_5  \\
           r_4s_1t_1-\bar{r}_3s_2t_5 
        \end{array}\right].
\end{align}
We find the Berry connection associated with this branch to be  
\begin{align}\label{eq.B-conn-p}
{\cal\vec{A}}^+_{\vec{q}} = 2DJa_0\sin^3\xi\left(\frac{W_1\left(A_{\vec{q}}^{11}- A_{\vec{q}}^{22}\right)S^2}{E_+(W_1-W_2)\left|A_{\vec{q}}^{12}\right|^2}-\frac{S^2}{\left|A_{\vec{q}}^{12}\right|^2}\right)\left[\begin{array}{c}\sin q_ya_0\\
\sin q_xa_0 \end{array}\right].
\end{align}
The lower magnon branch (i.e., with spin-wave energy $E_-$) is the second column of $\mat{T}_{\vec{q}}$,  
\begin{align}\label{eq:Psi-m}
|\Psi_{\vec{q}}^-\rangle =  \left[\begin{array}{c}
       r_1s_2t_3 + r_2s_1t_7 \\
       r_3s_1t_3 + r_4s_2t_7\\
          \bar{r}_2s_1t_3+r_1s_2t_7 \\
           r_4s_2t_3+\bar{r}_3s_1t_7 
        \end{array}\right].
\end{align}
This branch similarly corresponds to the Berry connection
\begin{align}\label{eq.B-conn-m}
{\cal\vec{A}}^-_{\vec{q}} = 2DJa_0\sin^3\xi\left(\frac{W_2\left(A_{\vec{q}}^{22}- A_{\vec{q}}^{11}\right)S^2}{E_-(W_1-W_2)\left|A_{\vec{q}}^{12}\right|^2}-\frac{S^2}{\left|A_{\vec{q}}^{12}\right|^2}\right)\left[\begin{array}{c}\sin q_ya_0\\
\sin q_xa_0 \end{array}\right].
\end{align}

\section{Spin-wave dynamics in the framework of the classical Landau-Lifshitz formalism}\label{app:LL}

In the classical limit, the dynamics of the magnetic subsystem can be described either by means of classical spins $\vec{S}_{\vec{r},1}(t)$, $\vec{S}_{\vec{r}',2}(t)$ or by means of the magnetic moments $\vec{\mu}_{\vec{r},1} \equiv g\mu_\textsc{b}\vec{S}_{\vec{r},1}$, $\vec{\mu}_{\vec{r}',2} \equiv g\mu_\textsc{b}\vec{S}_{\vec{r}',2}$. Let us first consider the spin representation. In this case, the dynamics obeys the following Landau-Lifshitz equation:
\begin{equation}\label{eq:LL-S}
    \dot{\vec{s}}_{\vec{\rho},\alpha}=-\frac{1}{\hbar S}\vec{s}_{\vec{\rho},\alpha}\times\frac{\partial\mathcal{H}}{\partial\vec{s}_{\vec{\rho},\alpha}},
\end{equation}
 where $\vec{s}_{\vec{\rho},\alpha}=\vec{S}_{\vec{\rho},\alpha}/S$ is the unit vector in the spin direction, $\alpha=1,2$ enumerates the sublattices, $\vec{\rho}=\vec{r}$ and $\vec{\rho}=\vec{r}'$ for $\alpha=1$ and $\alpha=2$, respectively. In the following, we utilize the classical analog of the Holstein-Primakoff representation~\eqref{eq.HP}
\begin{equation}\label{eq:HP-S}
    \vec{s}_{\vec{\rho},\alpha}=\vec{s}^0_\alpha\left(1-\frac{|\psi_{\vec{\rho},\alpha}|^2}{S}\right)+\sqrt{2-\frac{|\psi_{\vec{\rho},\alpha}|^2}{S}}\left(\vec{e}^-_\alpha\psi_{\vec{\rho},\alpha}+\vec{e}^+_\alpha\bar{\psi}_{\vec{\rho},\alpha}\right),
\end{equation}
where the complex-valued function $\psi_{\vec{\rho},\alpha}$ encodes deviation of the normalized spin $\vec{s}_{\vec{\rho},\alpha}$ from its equilibrium orientation $\vec{s}^0_{\alpha}$. Here $\vec{e}^\pm_\alpha=(\vec{e}^1_\alpha\pm i\vec{e}^2_\alpha)/(2\sqrt{S})$ with $\vec{e}^1_\alpha=\partial\vec{s}^0_\alpha/\partial\theta_\alpha$ and $\vec{e}^2_\alpha=\vec{s}^0_\alpha\times\vec{e}^1_\alpha$. Vectors $\vec{e}^\pm_\alpha$ have the following useful properties $\vec{e}^\pm_\alpha\cdot\vec{e}^\pm_\alpha=0$, $\vec{e}^\pm_\alpha\cdot\vec{e}^\mp_\alpha=1/2$, $\vec{e}^\pm_\alpha\cdot\vec{s}^0_\alpha=0$. In terms of $\psi_{\vec{\rho},\alpha}$, Landau-Lifshitz equations~\eqref{eq:LL-S} obtain Schr{\"o}dinger-like form 
\begin{equation}\label{eq:LL-psi}
    i\hbar\dot{\psi}_{\vec{\rho},\alpha}=\frac{\partial\mathcal{H}}{\partial \bar{\psi}_{\vec{\rho},\alpha}}.
\end{equation}
Applying the Fourier transform \eqref{eq.FT} to the linearized (with respect to $\psi_{\vec{\rho},\alpha}$) equation \eqref{eq:LL-psi}, we obtain
\begin{subequations}\label{eq:LL-psi-q}
\begin{align}    &i\hbar\dot{\psi}_{\vec{q},\alpha}=\frac{\partial\mathcal{H}^{\mathrm{cl}}_{2}}{\partial \bar{\psi}_{\vec{q},\alpha}},\\
&\mathcal{H}^{\mathrm{cl}}_{2}=\frac12\sum\limits_{\vec{q}}\vec{\Psi}^\dagger_{\vec{q}}\vec{M}_{\vec{q}}\vec{\Psi}_{\vec{q}},
\end{align}
\end{subequations}
where $\mathcal{H}^{\mathrm{cl}}_{2}$ is the harmonic part of the classical Hamiltonian. Here $\vec{\Psi}_{\vec{q}}=[\psi_{\vec{q},1},\psi_{\vec{q},2},\bar{\psi}_{-\vec{q},1},\bar{\psi}_{-\vec{q},2}]^{\intercal}$ and matrix $\vec{M}_{\vec{q}}$ coincides with one defined in \eqref{eq.M}. 
In terms of $\vec{\Psi}_{\vec{q}}$, equations of motion \eqref{eq:LL-psi-q} obtain the form of Schr{\"o}dinger equation 
\begin{equation}\label{eq:Psi}
    i\hbar\dot{\vec{\Psi}}_{\vec{q}}=\vec{H}_{\vec{q}}\vec{\Psi}_{\vec{q}}
\end{equation}
where $\vec{H}_{\vec{q}}=\vec{\eta}\vec{M}_{\vec{q}}$ with $\vec{\eta}$ defined in \eqref{eq.eta}. Although the matrix $\vec{\eta}\vec{M}_{\vec{q}}$ is not Hermitian, operator $\vec{H}_{\vec{q}}$ is Hermitian in the Hilbert space with pseudo-Euclidean metric $\vec{\eta}$.  The solution in form $\vec{\Psi}_{\vec{q}}=\vec{\Psi}^0_{\vec{q}}e^{-i\frac{E}{\hbar}t}$ reduces \eqref{eq:Psi} to the eigenvalue problem
\begin{equation}\label{eq:evp}
    E\vec{\Psi}^0_{\vec{q}}=\vec{H}_{\vec{q}}\vec{\Psi}^0_{\vec{q}},
\end{equation}
 which coincides with the eigenvalue problem in Eq.~\eqref{eq.EVP-Q}.  The eigenvalues $(E_{+}, E_{-}, -E_{+}, -E_{-})$ and the corresponding eigenvectors are discussed in sections \ref{sbsc:diagonal}, and \ref{sbsc:BC}. If the spin wave with wave-vector $\vec{q}$ is excited, then the real-space dynamics of spins in each of the sublattices is as follows
\begin{equation}\label{eq:s-reconstr}
    \begin{split}
&\vec{s}^{\nu}_{\vec{r},1}\approx\vec{s}^{0}_{1}+\sqrt{\frac{2}{N}}\mathrm{Re}\Biggl[e^{i(\vec{q}\cdot\vec{r}-\frac{E_\nu}{\hbar}t+\varphi_0)}\left(\vec{e}^-_1|\Psi^\nu_{\vec{q}}\rangle_1+\vec{e}^+_1|\Psi^\nu_{\vec{q}}\rangle_3\right)\Biggr],\\
&\vec{s}^{\nu}_{\vec{r}',2}\approx\vec{s}^{0}_{2}+\!\sqrt{\frac{2}{N}}\mathrm{Re}\Biggl[e^{i(\vec{q}\cdot\vec{r}'-\frac{E_\nu}{\hbar}t+\varphi_0)}\left(\vec{e}^-_2|\Psi^\nu_{\vec{q}}\rangle_2+\vec{e}^+_2|\Psi^\nu_{\vec{q}}\rangle_4\right)\Biggr],
\end{split}
\end{equation}
where $\nu=\pm$ denotes the magnon branch, $\varphi_0$ is an arbitrary phase, and $|\Psi^\nu_{\vec{q}}\rangle_{1,2,3,4}$ are  the components (rows) of vectors \eqref{eq:Psi-p} and \eqref{eq:Psi-m}.

To describe the dynamics of the set of magnetic moments, we introduce the unit magnetic moment $\vec{m}_{\vec{\rho},\alpha}=\vec{\mu}_{\vec{\rho},\alpha}/|\vec{\mu}_{\vec{\rho},\alpha}|$. The case $g>0$ coincides with the one considered above up to the interchange of the notations $\vec{s}_{\vec{\rho},\alpha}\leftrightarrow\vec{m}_{\vec{\rho},\alpha}$. However, for the case $g<0$, the corresponding Landau-Lifshitz equation for $\vec{m}_{\vec{\rho},\alpha}$ differ from \eqref{eq:LL-S} by sign:
\begin{equation}\label{eq:LL-m}
    \dot{\vec{m}}_{\vec{\rho},\alpha}=\frac{1}{\hbar S}\vec{m}_{\vec{\rho},\alpha}\times\frac{\partial\mathcal{H}}{\partial\vec{m}_{\vec{\rho},\alpha}}.
\end{equation}
Using the Holstein-Primakoff representation in the form 
\begin{equation}\label{eq:HP-m}
    \vec{m}_{\vec{\rho},\alpha}=\vec{m}^0_\alpha\left(1-\frac{|\psi_{\vec{\rho},\alpha}|^2}{S}\right)+\sqrt{2-\frac{|\psi_{\vec{\rho},\alpha}|^2}{S}}\left(\vec{e}^-_\alpha\bar{\psi}_{\vec{\rho},\alpha}+\vec{e}^+_\alpha\psi_{\vec{\rho},\alpha}\right),
\end{equation}
with $\vec{e}^1_\alpha=\partial\vec{m}^0_\alpha/\partial\theta_\alpha$ and $\vec{e}^2_\alpha=\vec{m}^0_\alpha\times\vec{e}^1_\alpha$ we write \eqref{eq:LL-m} in the same form as \eqref{eq:LL-psi}. Note that in two representations \eqref{eq:HP-S} and \eqref{eq:HP-m} the functions $\psi_{\vec{\rho},\alpha}$ and $\bar{\psi}_{\vec{\rho},\alpha}$ are interchanged. The latter compensates the change of the sign in the Landau-Lifshitz equation \eqref{eq:LL-m} leading to the same linerized equation \eqref{eq:LL-psi-q} in which 
\begin{equation}
    \mathcal{H}^{\mathrm{cl}}_{2}=\frac12\sum\limits_{\vec{q}}\vec{\Psi}^\dagger_{\vec{q}}\tilde{\vec{M}}_{\vec{q}}\vec{\Psi}_{\vec{q}}.
\end{equation}
The structure of the vectors $\vec{\Psi}_{\vec{q}}=[\psi_{\vec{q},1},\psi_{\vec{q},2},\bar{\psi}_{-\vec{q},1},\bar{\psi}_{-\vec{q},2}]^{\intercal}$ is the same as previously and $\tilde{\vec{M}}_{\vec{q}}=\left.\bar{\vec{M}}_{\vec{q}}\right|_{\xi\to-\xi}$. Note that the ground states for $\vec{m}_{\vec{\rho},\alpha}$ and $\vec{s}_{\vec{\rho},\alpha}$ are different if $g<0$. In the case under consideration, $\tilde{\vec{M}}_{\vec{q}}=\vec{M}_{\vec{q}}$. Thus, the equation of motion in $\vec{q}$-space and the corresponding eigenvalue problem coincide with \eqref{eq:Psi} and \eqref{eq:evp}, respectively. The latter means that the Berry curvatures computed from the $\vec{m}_{\vec{\rho},\alpha}$ and $\vec{s}_{\vec{\rho},\alpha}$ representations are the same.

In real space, the dynamics of the magnetic moments $\vec{m}_{\vec{\rho},\alpha}$ in a spin wave with wave-vector $\vec{q}$ can be reconstructed from \eqref{eq:HP-m} as follows:
\begin{align}\label{eq:m-reconstr}
&\vec{m}^{\nu}_{\vec{r},1}\!\approx\!\vec{m}^{0}_{1}+\!\sqrt{\frac{2}{N}}\mathrm{Re}\Biggl[e^{i(\vec{q}\cdot\vec{r}-\frac{E_\nu}{\hbar}t+\tilde{\varphi}_0)}\left(\vec{e}^+_1|\Psi^\nu_{\vec{q}}\rangle_1+\vec{e}^-_1|\Psi^\nu_{\vec{q}}\rangle_3\right)\Biggr],\nonumber\\
&\vec{m}^{\nu}_{\vec{r}',2}\!\approx\!\vec{m}^{0}_{2}+\!\sqrt{\frac{2}{N}}\mathrm{Re}\Biggl[e^{i(\vec{q}\cdot\vec{r}'-\frac{E_\nu}{\hbar}t+\tilde{\varphi}_0)}\left(\vec{e}^+_2|\Psi^\nu_{\vec{q}}\rangle_2+\vec{e}^-_2|\Psi^\nu_{\vec{q}}\rangle_4\right)\Biggr].
\end{align}
In \eqref{eq:m-reconstr}, one can use the expression for the eigenvectors derived in Eqs.~\eqref{eq:Psi-p} and \eqref{eq:Psi-m} with the replacement $|\Psi^\nu_{\vec{q}}\rangle_n\to|\bar{\Psi}^\nu_{\vec{q}}\rangle_n$, since Eqs.~\eqref{eq:Psi-p} and \eqref{eq:Psi-m} are obtained for $g>0$.

\section{Altermagnetic limit}\label{app:H0}
\begin{figure}
\centering
\includegraphics[width=0.9\columnwidth]{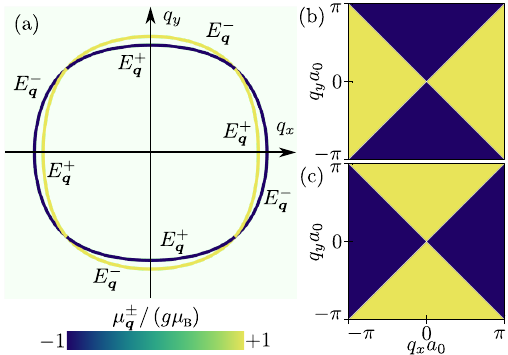}
\caption{(a) Distribution of the amplitude of the magnon magnetic moment~\eqref{eq:magnetic-moment} along the isolines of constant energy $E_{\bm{q}}^\pm=\text{const}$ in the limit of the Zeeman field $H=0$ and for the AM coupling $J_1 = -0.04J$. Panels (b) and (c) show the distribution of magnon magnetic moment~\eqref{eq:magnetic-moment} within the Brillouin zone plotted. Panels (b) and (c) correspond to the upper and lower branches, respectively.} 
\label{fig.magnetic_moment}
\end{figure}

In the limit $H=0$, the expression for the eigenenergies in Eq.~\eqref{eq:Epm} is reduced to
\begin{equation}
    E^\pm_{\vec{q}}=\frac12\left[\sqrt{\left(A^{11}_{\vec{q}}+A^{22}_{\vec{q}}\right)^2-4\left(B^{12}_{\vec{q}}\right)^2}\pm\left|A^{11}_{\vec{q}}-A^{22}_{\vec{q}}\right|\right].
\end{equation}
The corresponding magnon magnetic moment is
\begin{equation}\label{eq:magnetic-moment}
    \vec{\mu}^\pm_{\vec{q}}=\mp g\mu_B\,\text{sgn}\!\left(A^{11}_{\vec{q}}-A^{22}_{\vec{q}}\right)\vec{n}_0,
\end{equation}
where $\vec{n}_0=(\vec{S}_1-\vec{S}_2)/(2S)$ is the ground state staggered magnetization. The distribution of $\vec{\mu}^\pm_{\vec{q}}$ along the iso-lines $E^\pm_{\vec{q}}=\text{const}$ and within the Brillouin zone are shown in Fig.~\subref{fig.magnetic_moment}{(a)} and \subref{fig.magnetic_moment}{(b,c)}, respectively.

%%%%%%%%%%%%%%%%%%%%%%%%%%%%%%%%%%
%%%%%%%%%%%%%%%%%%%%%%%%%%%%%%%%%%

\section{Edge states in the framework of the classical Landau-Lifshitz formalism}\label{app:LL-edge}

%%%%%%%%%%%%%%%%%%%%%%%%%%%%%%%%%%
\begin{figure}[b]
\centering
\includegraphics[width=\columnwidth]{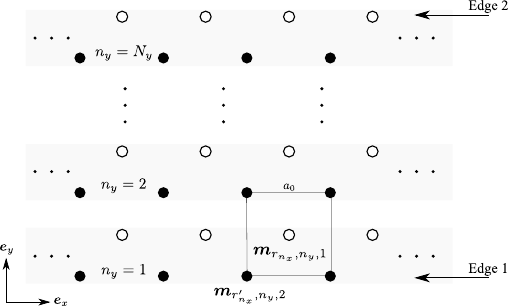}
\caption{The checkerboard strip which is infinite in the $\vhat{x}$-direction and contains $N_y$ spins of each of the sublattices in the $\vhat{y}$-direction. Index $n_y$ numerates shells shown by the gray shadowing. } 
\label{fig.stripe-edge}
\end{figure}
%%%%%%%%%%%%%%%%%%%%%%%%%%%%%%%%%%

Let's compute the spin waves in a strip which is infinite in $\vhat{x}$-direction and contains $N_y$ magnetic moments of each of the sublattices in the $\vhat{y}$-direction, see Fig.~\ref{fig.stripe-edge}. We base our analysis on the Landau-Lifshitz equation~\eqref{eq:LL-m} for magnetic moments where we assumed $g<0$. Each magnetic moment $\vec{m}_{r_{n_x},n_y,\nu}$ is uniquely identified by three quantities: (i) the sublattice $\nu=1,2$; (ii) position along the $\vhat{x}$-axis which is $r_{n_x}=a_0n_x$ and $r'_{n_x}=a_0n_x-\delta r$ for sublattices 1 and 2, respectively. Here $n_x\in\mathbb{Z}$, and $\delta r=a_0/2$; (iii) number $n_y=1,\dots,N_y$ of the shell along the $\vhat{y}$-direction.

Magnetic Hamiltonian of the strip $\mathcal{H}^{\mathrm{str}}=\mathcal{H}^{\mathrm{str}}_{\textsc{afm}}+\mathcal{H}^{\mathrm{str}}_{\textsc{alt}}+\mathcal{H}^{\mathrm{str}}_{\textsc{dmi}}+\mathcal{H}^{\mathrm{str}}_{\textsc{z}}$ comprises the antiferromagnetic
\begin{align}\label{eq:H-str-afm}
        \mathcal{H}^{\mathrm{str}}_{\textsc{afm}}=J\sum\limits_{r_{n_x}}\Biggl[&\sum\limits_{n_y=1}^{N_y}\vec{m}_{r_{n_x},n_y,1}\cdot(\vec{m}_{r_{n_x}+\delta r,n_y,2}+\vec{m}_{r_{n_x}-\delta r,n_y,2})\nonumber\\
        +&\sum\limits_{n_y=1}^{N_y-1}\vec{m}_{r_{n_x},n_y,1}\cdot(\vec{m}_{r_{n_x}+\delta r,n_y+1,2}+\vec{m}_{r_{n_x}-\delta r,n_y+1,2})\Biggr],
\end{align}
altermagnetic
\begin{align}\label{eq:H-str-alt}
   %\begin{split}
        \nonumber\mathcal{H}^{\mathrm{str}}_{\textsc{alt}}=J_1\sum\limits_{r_{n_x}}\biggl(&\sum\limits_{n_y=1}^{N_y-1}\vec{m}_{r_{n_x},n_y,1}\cdot\vec{m}_{r_{n_x},n_y+1,1}\\
        +&\sum\limits_{n_y=1}^{N_y}\vec{m}_{r_{n_x}-\delta r,n_y,2}\cdot\vec{m}_{r_{n_x}+\delta r,n_y,2}\biggr),
    %\end{split}
\end{align}
DMI
\begin{align}\label{eq:H-str-dmi}
    \nonumber\mathcal{H}^{\mathrm{str}}_{\textsc{dmi}}=&D\hat{\vec{z}}\cdot\sum\limits_{r_{n_x}}\Biggl[\sum\limits_{n_y=1}^{N_y}\vec{m}_{r_{n_x},n_y,1}\times(\vec{m}_{r_{n_x}+\delta r,n_y,2}-\vec{m}_{r_{n_x}-\delta r,n_y,2})\\
    +&\sum\limits_{n_y=1}^{N_y-1}\vec{m}_{r_{n_x},n_y,1}\times(\vec{m}_{r_{n_x}-\delta r,n_y+1,2}-\vec{m}_{r_{n_x}+\delta r,n_y+1,2})\Biggr],
\end{align}
and Zeeman 
\begin{equation}\label{eq:H-str-zee}
   \mathcal{H}^{\mathrm{str}}_{\textsc{z}}=-\mu_0 |g| \mu_BS\vec{H}\cdot\sum\limits_{r_{n_x}}\sum\limits_{n_y=1}^{N_y}(\vec{m}_{r_{n_x},n_y,1}+\vec{m}_{r_{n_x}-\delta r,n_y,2})
\end{equation}
contributions with $\vec{H}=H\hat{\vec{z}}$. Using the Holstein-Primakoff representation~\eqref{eq:HP-m} and applying a one-dimensional Fourier transform 
\begin{equation}\label{eq:four-1D}
\begin{split}
    &\psi_{r_{n_x},n_y,\nu}=\frac{1}{\sqrt{N_x}}\sum\limits_{q_x}\psi_{q_x,n_y,\nu}e^{iq_xr_{n_x}},\\
    &\psi_{q_x,n_y,\nu}=\frac{1}{\sqrt{N_x}}\sum\limits_{r_{n_x}}\psi_{r_{n_x},n_y,\nu}e^{-iq_xr_{n_x}},
\end{split}
\end{equation}
where $q_x\in[-\pi/a_0,\pi/a_0]$, we obtain the equations of motion for $\vec{\Psi}_{q_x}=[\vec{\psi}_{q_x,1},\ldots,\vec{\psi}_{q_x,N_y}]^{\intercal}$ with $\vec{\psi}_{q_x,n_y}=[\psi_{q_x,n_y,1},\psi_{q_x,n_y,2},\bar{\psi}_{-q_x,n_y,1},\bar{\psi}_{-q_x,n_y,2}]^{\intercal}$ in the form of Schr{\"o}dinger equation
\begin{equation}\label{eq:Psi-edge}
    i\hbar\dot{\vec{\Psi}}_{q_x}=\vec{H}^{\mathrm{str}}_{q_x}\vec{\Psi}_{q_x},\qquad \vec{H}^{\mathrm{str}}_{q_x}=\left[\begin{array}{ccccccc}
         \vec{S}_1& \vec{R} & 0 & & \dots & & 0  \\
         \vec{L} & \vec{S} &\vec{R} & 0 & &\dots  & 0 \\
         0 & \vec{L} & \vec{S} &\vec{R} & 0 & \dots & 0 \\
         & & & & \ddots  & &\\
         0 & \dots& & 0& \vec{L} & \vec{S} &\vec{R} \\
         0 & &\dots & & 0& \vec{L} & \vec{S}_{2} 
    \end{array}
    \right],
\end{equation}
\begin{widetext}
where $\vec{H}^{\mathrm{str}}_{q_x}$ is $4N_y\times4N_y$ matrix and $\vec{S}_1$, $\vec{S}_{2}$, $\vec{S}$, $\vec{L}$, and $\vec{R}$ are the following block matrices
    \begin{equation}
        \begin{split}
            \vec{S}_1=&E_0\begin{bmatrix}
                1-\frac{J_1}{4J}&-\mathscr{A}_{q_x}+\mathscr{B}_{q_x}&0&\mathscr{C}_{q_x}\\
                -\mathscr{A}_{q_x}+\mathscr{B}_{q_x}&\mathscr{F}-\frac{J_1}{J}\sin^2\frac{q_xa_0}{2}&\mathscr{C}_{q_x}&0\\
                0&-\mathscr{C}_{q_x}&-1+\frac{J_1}{4J}&\mathscr{A}_{q_x}+\mathscr{B}_{q_x}\\
                -\mathscr{C}_{q_x}&0&\mathscr{A}_{q_x}+\mathscr{B}_{q_x}&-\mathscr{F}+\frac{J_1}{J}\sin^2\frac{q_xa_0}{2}
            \end{bmatrix},\quad\vec{R}=E_0\begin{bmatrix}
                \frac{J_1}{4J}&-\mathscr{A}_{q_x}-\mathscr{B}_{q_x}&0&\mathscr{C}_{q_x}\\
                0&0&0&0\\
                0&-\mathscr{C}_{q_x}&-\frac{J_1}{4J}&\mathscr{A}_{q_x}-\mathscr{B}_{q_x}\\
                0&0&0&0
            \end{bmatrix},\\
            \vec{S}=&E_0\begin{bmatrix}
                1-\frac{J_1}{2J}&-\mathscr{A}_{q_x}+\mathscr{B}_{q_x}&0&\mathscr{C}_{q_x}\\
                -\mathscr{A}_{q_x}+\mathscr{B}_{q_x}&1-\frac{J_1}{J}\sin^2\frac{q_xa_0}{2}&\mathscr{C}_{q_x}&0\\
                0&-\mathscr{C}_{q_x}&-1+\frac{J_1}{2J}&\mathscr{A}_{q_x}+\mathscr{B}_{q_x}\\
                -\mathscr{C}_{q_x}&0&\mathscr{A}_{q_x}+\mathscr{B}_{q_x}&-1+\frac{J_1}{J}\sin^2\frac{q_xa_0}{2}
            \end{bmatrix},\quad\vec{L}=E_0\begin{bmatrix}
                 \frac{J_1}{4J}&0&0&0\\
                -\mathscr{A}_{q_x}-\mathscr{B}_{q_x}&0&\mathscr{C}_{q_x}&0\\
                0&0&-\frac{J_1}{4J}&0\\
                -\mathscr{C}_{q_x}&0&\mathscr{A}_{q_x}-\mathscr{B}_{q_x}&0
            \end{bmatrix},\\
            \vec{S}_2=&E_0\begin{bmatrix}
                \mathscr{F}-\frac{J_1}{4J}&-\mathscr{A}_{q_x}+\mathscr{B}_{q_x}&0&\mathscr{C}_{q_x}\\
                -\mathscr{A}_{q_x}+\mathscr{B}_{q_x}&1-\frac{J_1}{J}\sin^2\frac{q_xa_0}{2}&\mathscr{C}_{q_x}&0\\
                0&-\mathscr{C}_{q_x}&-\mathscr{F}+\frac{J_1}{4J}&\mathscr{A}_{q_x}+\mathscr{B}_{q_x}\\
                -\mathscr{C}_{q_x}&0&\mathscr{A}_{q_x}+\mathscr{B}_{q_x}&-1+\frac{J_1}{J}\sin^2\frac{q_xa_0}{2}
            \end{bmatrix}.
        \end{split}
    \end{equation}
\end{widetext}
Here we introduced the following notations $\mathscr{A}_{q_x}=\frac12\cos\frac{q_xa_0}{2}\sin^2\xi$, $\mathscr{B}_{q_x}=\frac{D}{2J}\sin\frac{q_xa_0}{2}\sin\xi$, $\mathscr{C}_{q_x}=\frac12\cos\frac{q_xa_0}{2}\cos^2\xi$, and $\mathscr{F}=\frac{1}{2}+\sin^2\xi$. The solution of Eq.~\eqref{eq:Psi-edge} is $\vec{\Psi}_{q_x}=\vec{\Psi}^0_{q_x} e^{-i \frac{E_{q_x}}{\hbar} t}$, where $E_{q_x}$ is an eigenvalue of matrix $\vec{H}^{\mathrm{str}}_{q_x}$. An example of the spectrum generated by numerically diagonalizing the matrix $\vec{H}^{\mathrm{str}}_{q_x}$ is shown in Fig.~\ref{fig.edge}.
\bibliography{main}
\end{document}